\documentclass[prd,singlecolumn]{revtex4}
\pdfoutput=1
\usepackage{amssymb,latexsym}
\usepackage{amsmath,amsbsy,bbm}
\usepackage{ifpdf}
\usepackage{epsfig,bm}
\usepackage{graphicx,comment}
\usepackage{color}
\usepackage{soul}
\usepackage{mathtools}
\usepackage{comment}
\usepackage[normalem]{ulem}
\begin{document}

\title{Comments on the minimal training set for CNN: a case study of the frustrated $J_1$-$J_2$ Ising model on the square lattice }
\author{Shang-Wei Li}
\affiliation{Department of Physics, National Taiwan Normal University,
  88, Sec.4, Ting-Chou Rd., Taipei 116, Taiwan}
    \author{Yuan-Heng Tseng}
  \affiliation{Department of Physics, National Taiwan Normal University,
  	88, Sec.4, Ting-Chou Rd., Taipei 116, Taiwan}
  \author{Ming-Che Hsieh}
  \affiliation{Department of Physics, National Taiwan Normal University,
  	88, Sec.4, Ting-Chou Rd., Taipei 116, Taiwan}
  	  \author{Fu-Jiun Jiang}
\email[]{fjjiang@ntnu.edu.tw}
\affiliation{Department of Physics, National Taiwan Normal University,
88, Sec.4, Ting-Chou Rd., Taipei 116, Taiwan}

\begin{abstract}
The minimal training set to train a working CNN is explored in detail. The considered model is the frustrated $J_1$-$J_2$ Ising model on the square lattice. Here $J_1 < 0$ and $J_2 > 0$ are the nearest and next-to-nearest neighboring couplings, respectively. We train the CNN using the configurations of $g \stackrel{\text{def}}{=} J_2/|J_1| = 0.7$ and employ the resulting CNN to study the phase transition of
$g = 0.8$. We find that this transfer learning is successful. In particular, only configurations of two temperatures, one is below and one is above the critical temperature $T_c$ of $g=0.7$, are needed to reach accurately determination of the $T_c$ of $g=0.8$. However, it may be subtle to use this strategy for the training. Specifically, for the considered model, due to the inefficiency of the single spin flip algorithm used in sampling the configurations at the low-temperature region, the
two temperatures associated with the training set should not be too far away from the $T_c$ of $g=0.7$, otherwise, the performance of
the obtained CNN is not of high quality, hence cannot determine the $T_c$ of $g=0.8$ accurately. For the considered model, we also uncover the condition for training a successful CNN when only configurations of two temperatures are considered as the training set.

\end{abstract}

\maketitle

\section{Introduction}

Recently, both the supervised and unsupervised neural networks (NN) have emerged as powerful methods of achieving 
understanding the critical phenomena of physical systems \cite{Oht16,Tan16,Car16,Nie16,Meh19,Car19}. In particular, similar to the traditional ideas, these techniques can
calculate with fair precision of both the critical points as well as the associated exponents. Despite the fact that the broad adoption of NN in
exploring transitions between different phases took place only less than a decade ago, this modern approach paves a new route that may lead to understanding the critical phenomena from a different perspective.  

Conventionally, when supervised NN is used to study a phase transition, both the configurations from below and above the critical points 
are employed to train the supervised NN. To achieve a valid NN
that can determine the critical point with high accuracy,
one may need certain amount of data and certain number of epochs to train the supervised NN. This training procedure typically consumes 
a lot of computing resources and time. 
  
As a result, to advance the application of NN in investigating the critical phenomena, it is crucial and desirable to shorten 
the time for the training and to limit the use of computing resources.
A highly efficient NN should be obtained at the same time as well. 

There are some development in this regard. For instance, the transfer learning \cite{Cor21,Fuk21}, the use of theoretical ground states or artificially made
configurations instead of real configurations for the training \cite{Li18,Tan20,Tan20.1}, the consideration of fewer neurons and layers \cite{Kim18}, and so on.

Recently, the preprint of Ref.~\cite{Abu25} argues that
only the configurations of $T=0$ and $T=\infty$ are required
to train a successful supervised convolutional NN (CNN) to uncover
the critical theory of the Ising model on the square lattice. Here $T$
refers to the temperature. In Ref.~\cite{Abu25}, the configurations used
for the training in principle are the theoretical ground state configurations. 
For example, all the Ising spins of the configurations of $T=0$ used in the training stage take the same value of either 1 or -1. In addition, the configurations of $T=\infty$ are made random (in 1 and -1) manually. This approach is not
different from the strategy of using ground state configurations as the training set which is firstly introduced in Ref.~\cite{Li18}. Finally, in Ref.~\cite{Li18}, only
the configurations of $T=0$ are required and the data of $T=\infty$ are
not needed for the training.

In studying the phase transition associated with the temperature of a real system,
the data are determined experimentally within a temperature region. Therefore, if a phase transition is indeed observed in the considered temperature region of that experiment, then it will indeed be important and crucial to examine what is the minimal training set to obtain a supervised NN that is valid for investigating the targeted critical phenomenon. 

We would like to emphasize the fact again that for experiments of a system that no any information of that system is known, it is impossible to employ the related theoretical ground states as the training set to train any NN. Therefore, 
one can only rely on the real data to train a NN. Under such a circumstance,
it is indeed necessary to examine what is the minimal training set (based on the 
experimental data) so that a resulting NN can successfully explore the desired phase transition. 

One drawback of applying supervised NNs to study phase transitions is that 
the coarse location of the critical points should be available in advance before
one can perform the NN calculations. If it is generally true that only the configurations (experimental data) at the two ends of the considered parameter
are required to train a NN that works, then the rough position of the critical points is not necessary to conduct the supervised NN investigations. This will then put the supervised NNs at the same position of usefulness as their unsupervised counterparts.  

To further examine the validity of the training strategy of using only the configurations at the two ends of the considered parameter, here we investigate the phase transitions of the frustrated $J_1$-$J_2$ Ising model on the square lattice using three CNNs. Here $J_1 < 0$ and $J_2 > 0$ are the nearest and next-to-nearest neighboring couplings, respectively. This model has attracted
a lot of theoretical and numerical interests during the last several decades due to its potential applications in frustrated magnets \cite{Bin80,Lan80,Oit81,Lan85,Oit87,Lop931,Mal06,Kal08,Kal09,Jin12,Kal12,Jin13,Hu21,Li21,Yos23,Li24}.

We train these CNNs using the configurations of $g \stackrel{\text{def}}{=} J_2/|J_1| = 0.7$, and then employ the obtained CNNs to calculate the critical temperature of $g=0.8$.

Unexpectedly, although we do find only configurations of two temperatures, one is below and one is above the $T_c$ of $g=0.7$, are required to train a successful CNN that can calculate the $T_c$ of $g=0.8$ precisely, these two chosen temperatures cannot be too far away from the critical temperature of $g=0.7$, otherwise, the $T_c$ of $g=0.8$ is either not accurately determined or receives
fairly large finite-size effect(s).

By carefully examining the spin configurations in the training set(s),
we also uncover the condition for training a successful CNN when only
configurations of two temperatures are considered as the training set.

The rest of the paper is organized as follows. After the introduction which includes the motivation of current study, we detail the model and give a brief description of the used CNNs in Sect. II and Sect. III, respectively. Then the numerical outcomes are demonstrated thoroughly in Sect. IV. Finally, Sect. V contains our discussions and summaries of present investigation.

\section{The considered model}

\begin{figure}
    \includegraphics[width=0.3\textwidth]{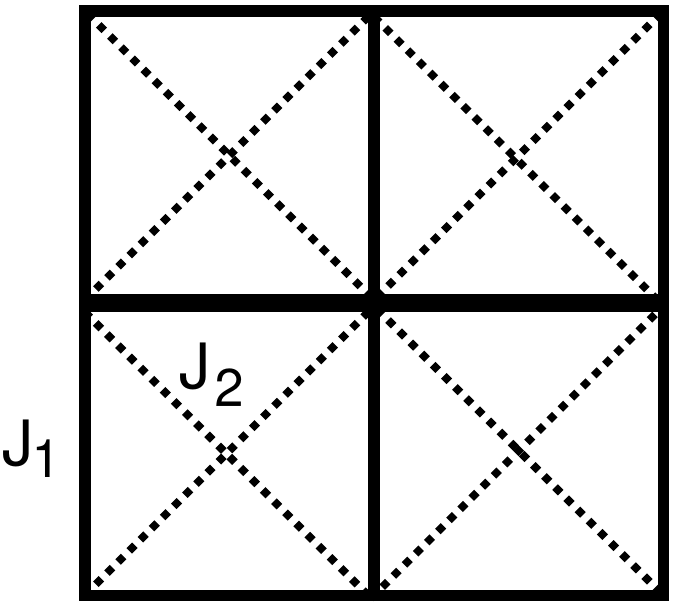}
        \caption{The frustrated $J_1$-$J_2$ Ising model on the square lattice studied here. The figure is taken from Ref.~\cite{Li24}.}
        \label{j1j2}
\end{figure}

The Hamiltonian $H$ of the frustrated $J_1$-$J_2$ Ising model on the square
lattice studied here has the following form
\begin{equation}
 H = J_1 \sum_{\left< ij\right>} \sigma_i\sigma_j + J_2 \sum_{\left<\left< lm\right>\right>}\sigma_l\sigma_m,
\label{eqn}
\end{equation}
where $\sigma_i = \pm 1$ is the Ising spin at site $i$. The first
and the second terms appearing in Eq.~(\ref{eqn}) are summed over nearest neighboring sites $i$ and $j$ and summed over next-to-nearest neighboring sites $l$ and $m$, respectively.
Fig.~\ref{j1j2} is the graphical representation of the model studied here. 
In our calculations, $J_1 = -1$ and $g \stackrel{\text{def}}{=} J_2/|J_1|$. The phase
transitions of $g = 0.7$ and $g=0.8$ are investigated in this study.

\section{The employed CNNs}

To ensure the claims found from the NN calculations do not depend on a
specific architecture of the used CNN,
in this study, we employ 3 different CNNs to investigate the proposed phase transitions. These CNNs are implemented using tensorflow and keras \cite{tens}, and their architectures are introduced briefly
in the following.

\subsection{The first CNN}
\begin{figure}
	\includegraphics[width=0.8\textwidth]{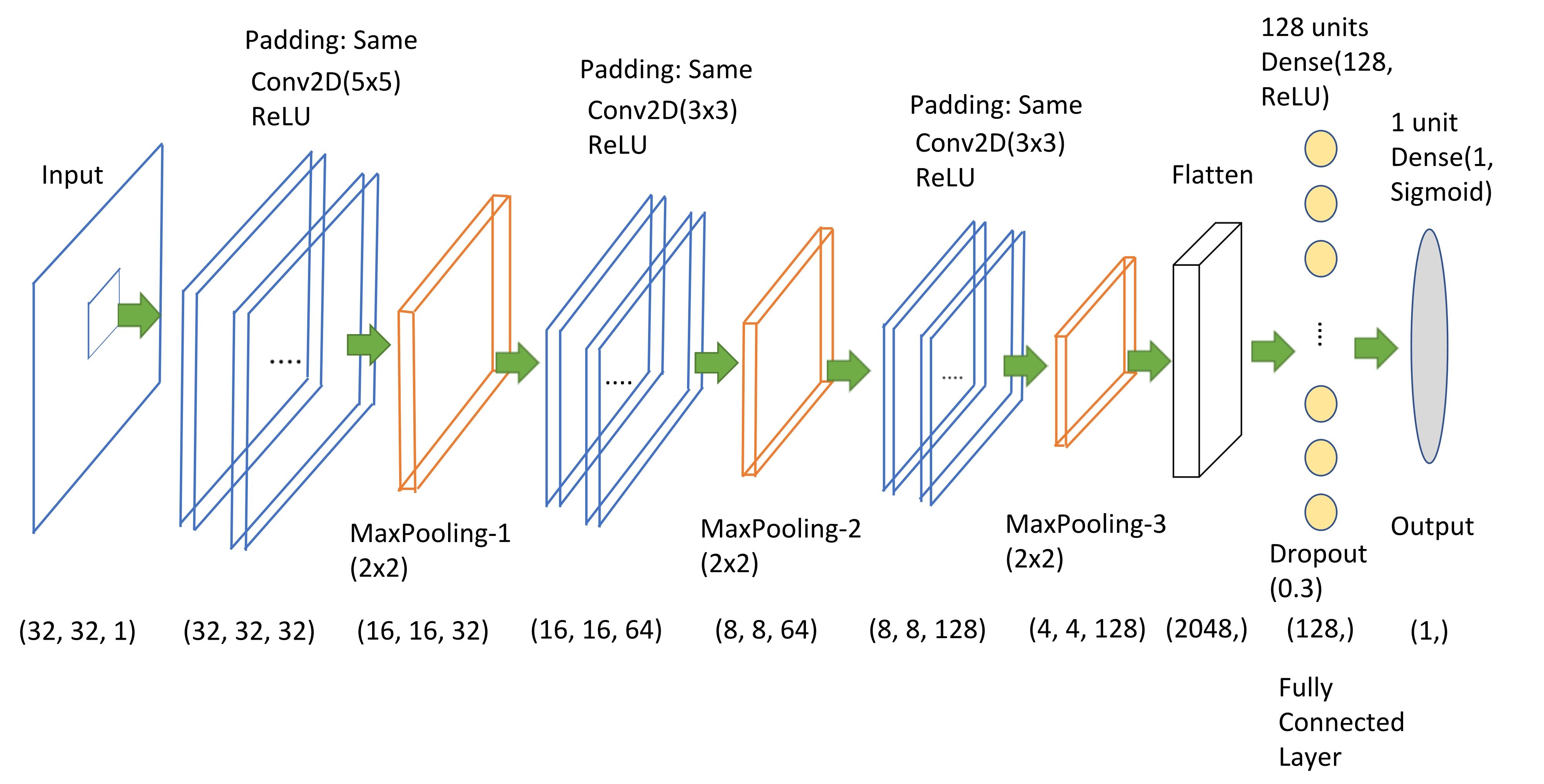}
	\caption{The first CNN used here. The numbers at the bottom of the figure are for $L=32$. The figure is created based on a figure of the submitted version of Ref.~\cite{Tse24}.}
	\label{cnn_picture0}
\end{figure}

The architecture of the first CNN employed to study the targeted phase transition(s)
is shown in fig.~\ref{cnn_picture0}. The employed activation functions, the shape of kernels, the padding scheme, the pooling method, and other CNN-related parameters are explicitly shown in the figure. The epochs considered is 10. Moreover, the loss function used is the BinaryEntropy. Finally, the CNN outputs are real numbers from 0 to 1, and 0 and 1 stand for the likeness that the associated configurations are in the ordered and disordered phases, respectively. The training
is conducted with configurations of $g=0.7$ and the resulting CNN is considered to
study the phase transition of $g=0.8$. This first CNN is named CNN1 for convenience.

\subsection{The second CNN}

\begin{figure}
		\includegraphics[width=0.85\textwidth]{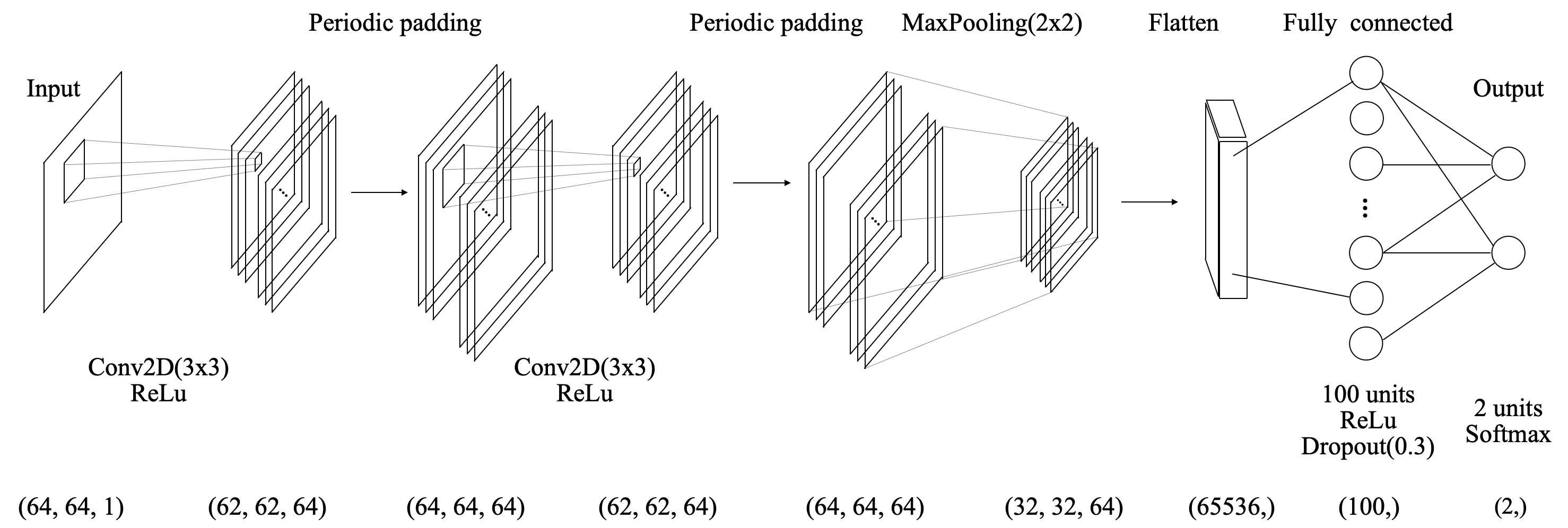}
	\caption{The second CNN used here. The figure is taken from the submitted version of Ref.~\cite{Tse24}. The numbers at the bottom of the figure are for $L=64$.}
	\label{cnn_picture}
\end{figure}

The second CNN used here is adopted from Ref.~\cite{Tse24} and the associated pictorial representation of its architecture is depicted in fig.~\ref{cnn_picture} (The architecture of the CNN of Ref.~\cite{Tse24} is based on that of Ref.~\cite{Tol23}). 
Moreover, the employed activation functions, the shape of kernels, the padding scheme, the pooling method, and other CNN-related parameters are explicitly shown in the figure. The CNN outputs are two-component vectors.
The first and the second component of the CNN output stand for the likeness that the input configuration are in the ordered and disordered phases, respectively.
This second CNN is named CNN2 for convenience.

\begin{figure}
	\hbox{~~~~~~~~~~~~                           
		\includegraphics[width=0.4\textwidth]{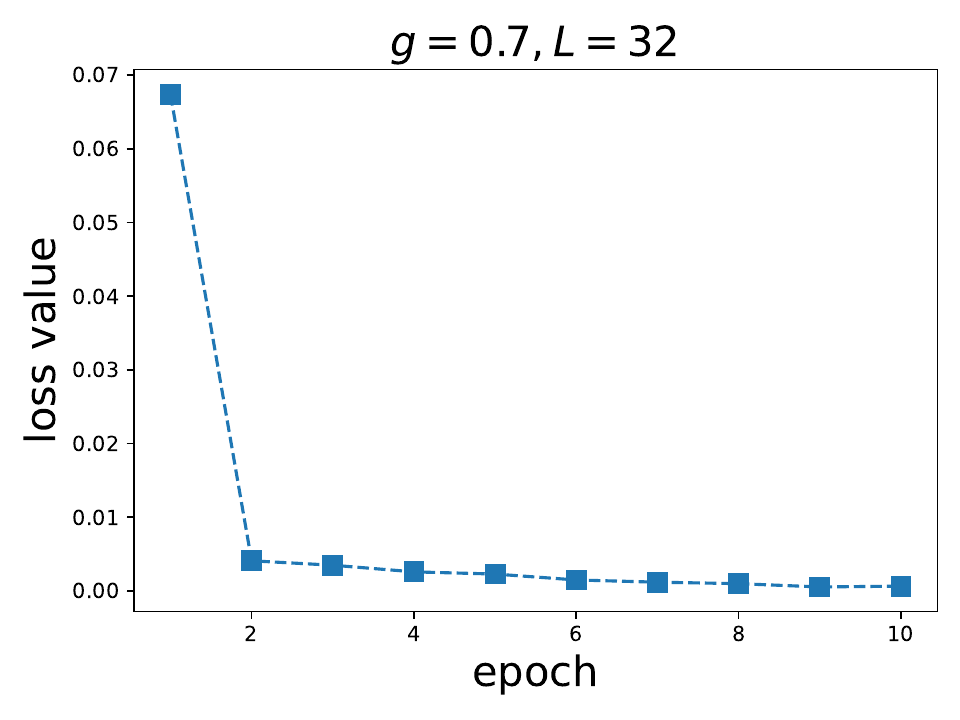}~~~~~~~~~~
		\includegraphics[width=0.4\textwidth]{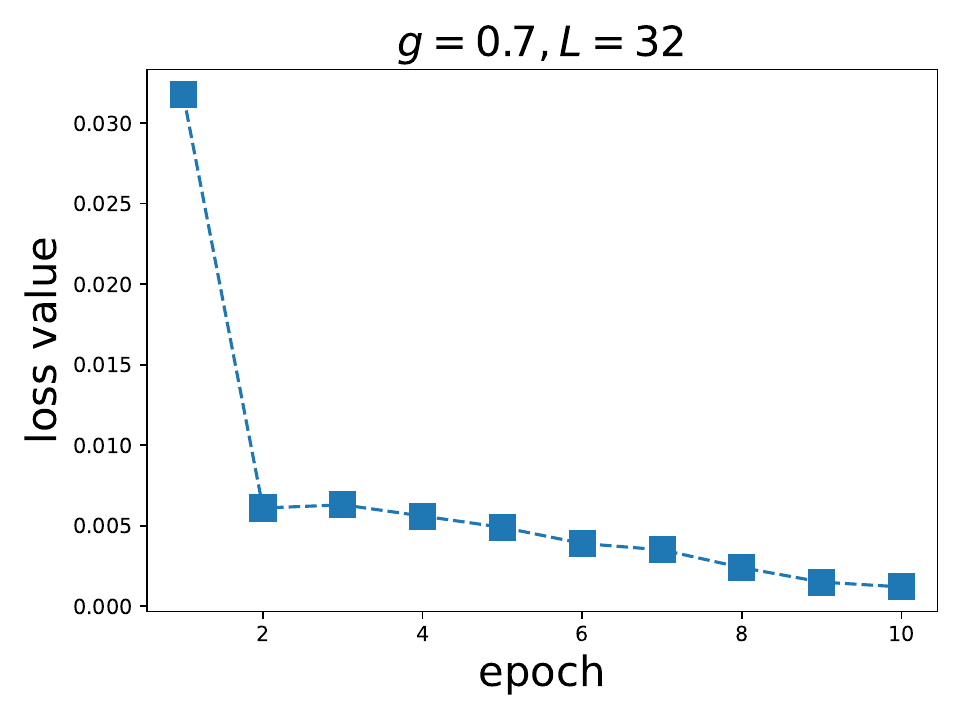}
	}                    
	
	\caption{The values of loss function during the training for conducting 10 epochs using CNN2. }
	
	\label{loss}
\end{figure}

Similar to CNN1, we use the configurations obtained with $g=0.7$ to train CNN2. 
During the training, several sets of random seeds are used and 4 to 20 epochs are conducted. The typical values of loss function (we use CategoryEntropy as the loss function)
during the training of performing 10 epochs are shown in fig.~\ref{loss}. The resulting CNN (CNN2) is then applied to study the phase transition of $g=0.8$.

\subsection{The third CNN}

The third CNN (which is denoted by CNN3 for convenience) considered here is two copies of CNN2.
Specifically, the third CNN has two copies of the convolutional layers and the fully connected layer shown in fig.~\ref{cnn_picture}. In summary, CNN3 has four convolutional layers and two fully connected layers. The associated CNN3 outputs
are two-component vectors like those of CNN2.

\section{Numerical Results}

To study the proposed transfer learning and to investigate what is the minimal training set to train a CNN that works, for each of the chosen temperatures for both $g=0.7$ and $g=0.8$, we have generated few to several thousand configurations using single spin flip Metropolis algorithm \cite{Met53}. It should be pointed out that for temperatures far below the critical temperature $T_c$, while one can reach one of the equilibrium steady ground states by the MC simulations, sampling among these
states is difficult because the associated MC chain is generated by a local 
update algorithm. One notices this limitation is applied to both the cases of $g=0.7$ and $g=0.8$.
Therefore, if the associated configurations obtained from the simulations at the low-temperature region of $g=0.7$ are considered in the training set(s), then apparently
the resulting CNN can identify the configurations of the low-$T$ region of $g=0.8$.
Evidence supporting this fact will be demonstrated later when the CNN outcomes 
are presented. As a result, both the low-$T$ configurations of
$g=0.7$ and $g=0.8$ are included in our investigation.

We would like to emphasize the fact again that the training is performed using the configurations
of $g=0.7$ and the resulting trained CNN is used to study the
phase transition associated with $g=0.8$. Finally, it should be pointed out that the $T_c$ of $g=0.7$ and $g=0.8$ are given by
1.2890 and 1.5678, respectively \cite{Jin13,Yos23,Li24}.

\subsection{The outcomes related to CNN1}

\begin{figure}
	\hbox{                           
		\includegraphics[width=0.45\textwidth]{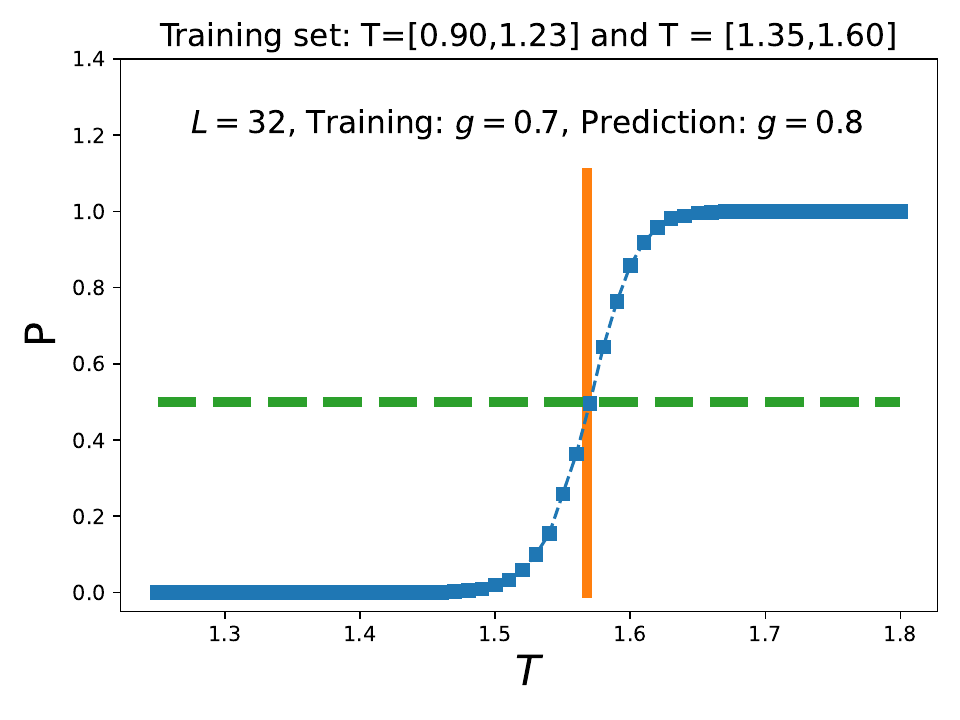}~~~~~~
		\includegraphics[width=0.45\textwidth]{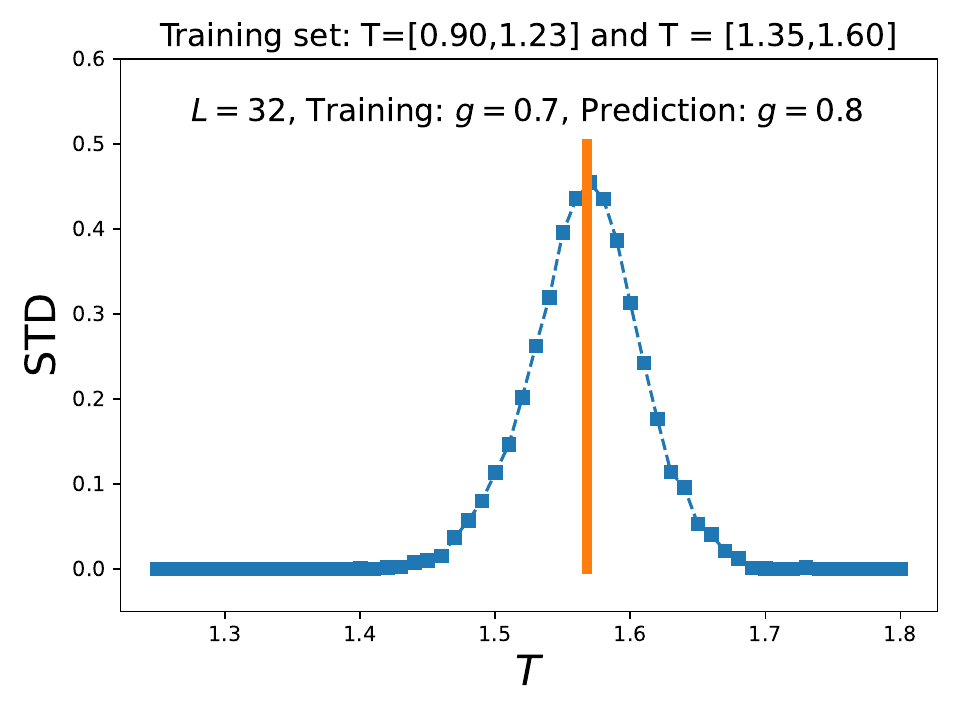}
	}                    
	
	\caption{The transfer learning results. 
		(Left) The NN outputs as a function of $T$. The horizontal dashed line is 0.5. (Right) The associated standard deviations (STD) as a function of $T$.
		In both panels, the vertical solid lines are the theoretical $T_c$ of $g=0.8$. These outcomes are associated with CNN1.}	
	\label{cnn01}
\end{figure}

The left panel of fig.~\ref{cnn01} demonstrates the outputs of CNN1 as a function
of $T$ for $g=0.8$ with $L=32$. The title shows the temperatures considered for
the training. In the training set, we excludes those configurations which the associated temperatures are near $T_c$.
In the figure, the horizontal dashed and vertical solid lines represent 0.5 and
the $T_c$ of $g=0.8$. The intersection of the outputs and 0.5 is the NN-determined critical temperature. As can be seen from the panel, the NN-determined critical temperature agrees very well with the theoretical $T_c$.

The STD of the NN outputs as a function of $T$ is depicted in the right panel of fig.~\ref{cnn01}. The panel reveals the message that at $T_c$, the magnitude of the STD of NN output is the largest.  

To summarize, both the NN outputs and the associated STD can be employed to
calculate the NN-related $T_c$ with fairly high accuracy. 

Despite the fact that the configurations obtained at low-$T$ region may
not have all the degenerated stripe states (due to the inefficiency of the single spin flip algorithm used here), the outcomes demonstrated in fig.~\ref{cnn01} provide evidence supporting the effectiveness of including these low-$T$ configurations in the CNN calculations. 

\begin{figure}
	\includegraphics[width=0.55\textwidth]{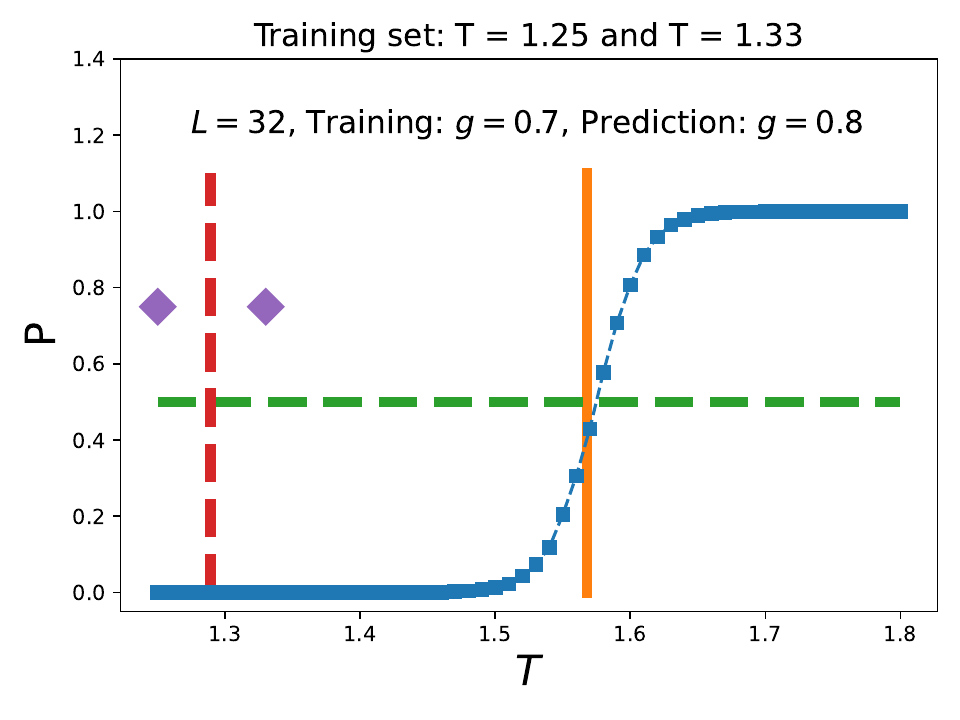}
	\caption{The transfer learning results. The CNN1 outputs as a function of $T$.
		The horizontal dashed line is 0.5. The training set consists of configurations of $T=1.25$ and $T=1.33$ (The diamond symbols). The vertical dashed and solid lines are the theoretical $T_c$ of $g=0.7$ and $g=0.8$, respectively. These outcomes are associated with CNN1}
	\label{cnn02}
\end{figure}

Fig.~\ref{cnn02} shows the CNN1 outputs as a function of $T$ for $g=0.8$ when
the configurations of $T=1.25$ and $T=1.33$ of $g=0.7$ are used as the training set. Moreover, the diamond symbols in the figure represent these two temperatures and the vertical dashed line is the $T_c$ of $g=0.7$. The vertical solid line is the expected $T_c$ of $g=0.8$. The results of the figure demonstrate that when the training set consists of
merely configurations from two temperatures, the obtained CNN1 can successfully
determine the $T_c$ of $g=0.8$ precisely. 

\begin{figure}
	\hbox{
	\includegraphics[width=0.32\textwidth]{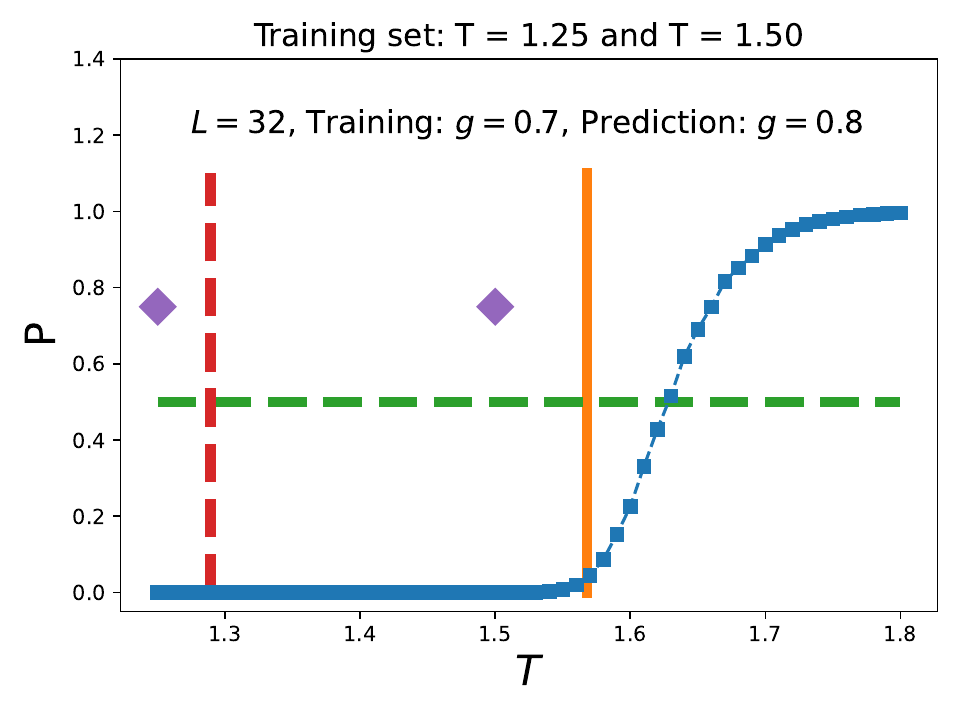}
\includegraphics[width=0.32\textwidth]{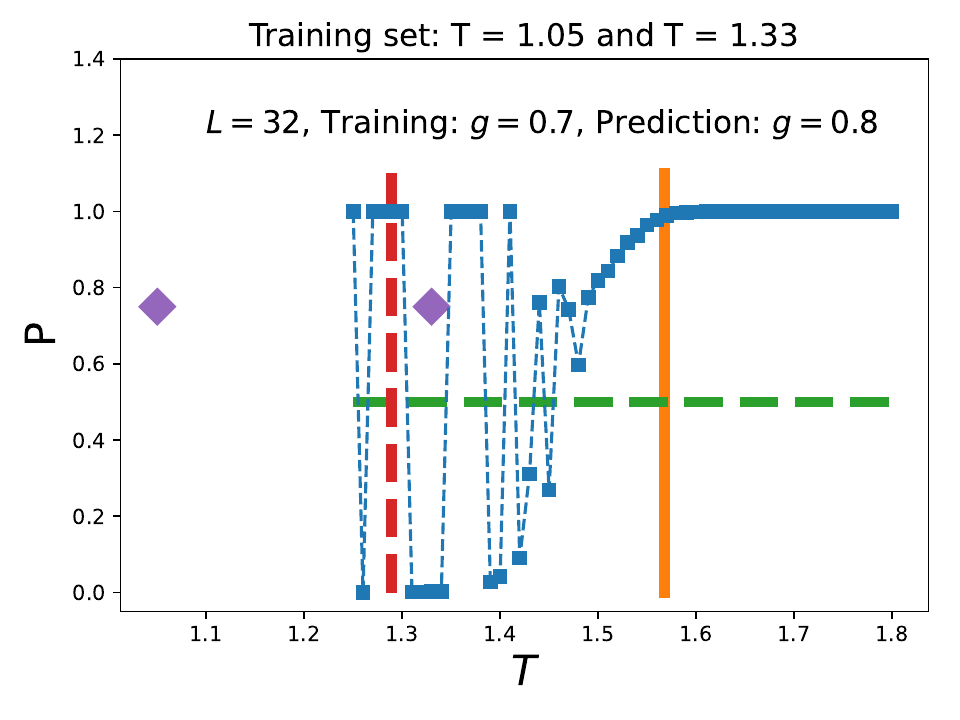}
\includegraphics[width=0.32\textwidth]{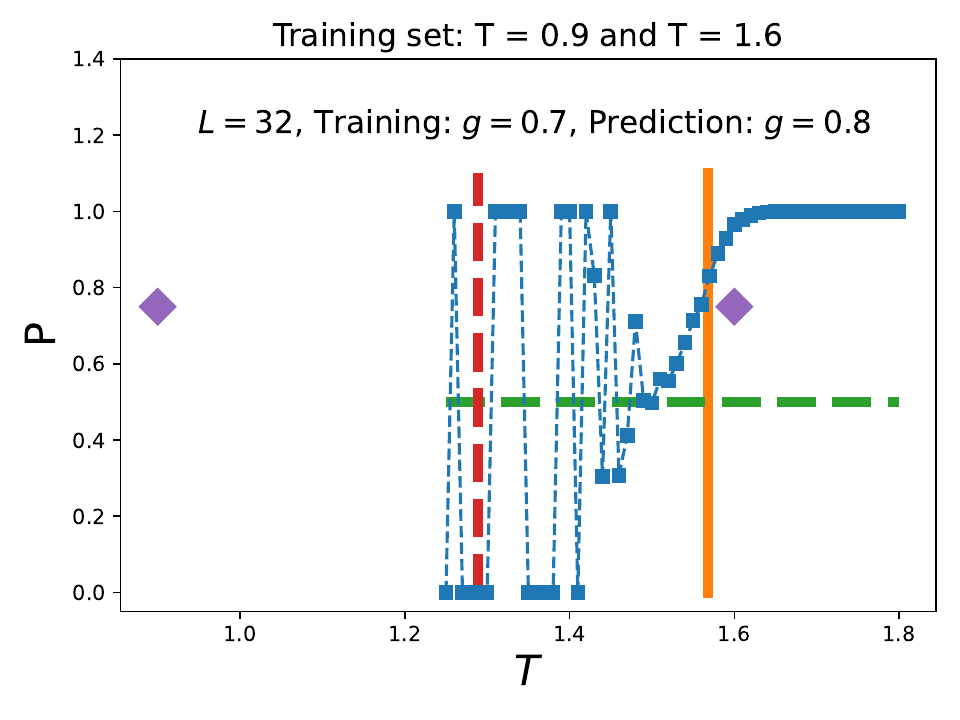}
}
	\caption{The transfer learning results. The CNN1 outputs as functions of $T$.
		The horizontal dashed lines are 0.5. Diamond symbols represent the temperatures for which the associated configurations are considered in the training sets. The vertical dashed and solid lines are the theoretical $T_c$ of $g=0.7$ and $g=0.8$, respectively. These outcomes are associated with CNN1.}
	\label{cnn03}
\end{figure}

Fig.~\ref{cnn03} shows the CNN1 outputs as functions of $T$ for $g=0.8$ when
the configurations of two temperatures of $g=0.7$ are used as the training sets
(The used temperatures are listed as the titles of the panels and are presented as the diamond symbols in the figure). The horizontal dashed, the vertical dashed,
and the vertical solid lines are 0.5, $T_c$ of $g=0.7$, and $T_c$ of $g=0.8$,
respectively.
As can be seen from all the panels of the figure, when one of (both) the temperatures is (are) far away from the $T_c$ of $g=0.7$, then one reaches poor determination of the $T_c$ of $g=0.8$ using the resulting CNN1. 

\begin{figure}
	\hbox{
		\includegraphics[width=0.45\textwidth]{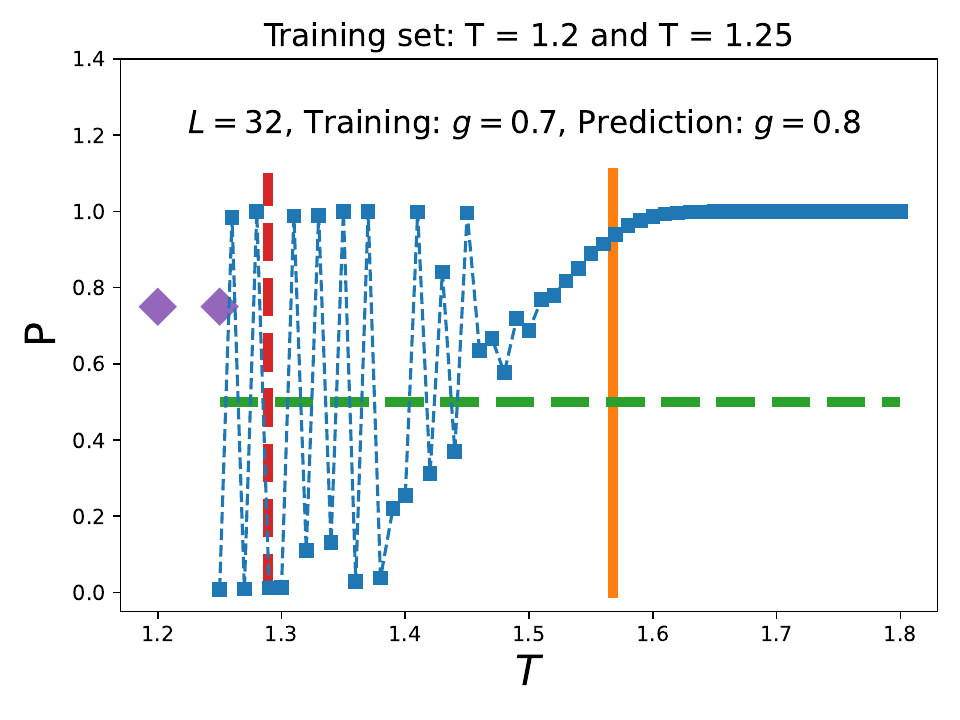}
		\includegraphics[width=0.45\textwidth]{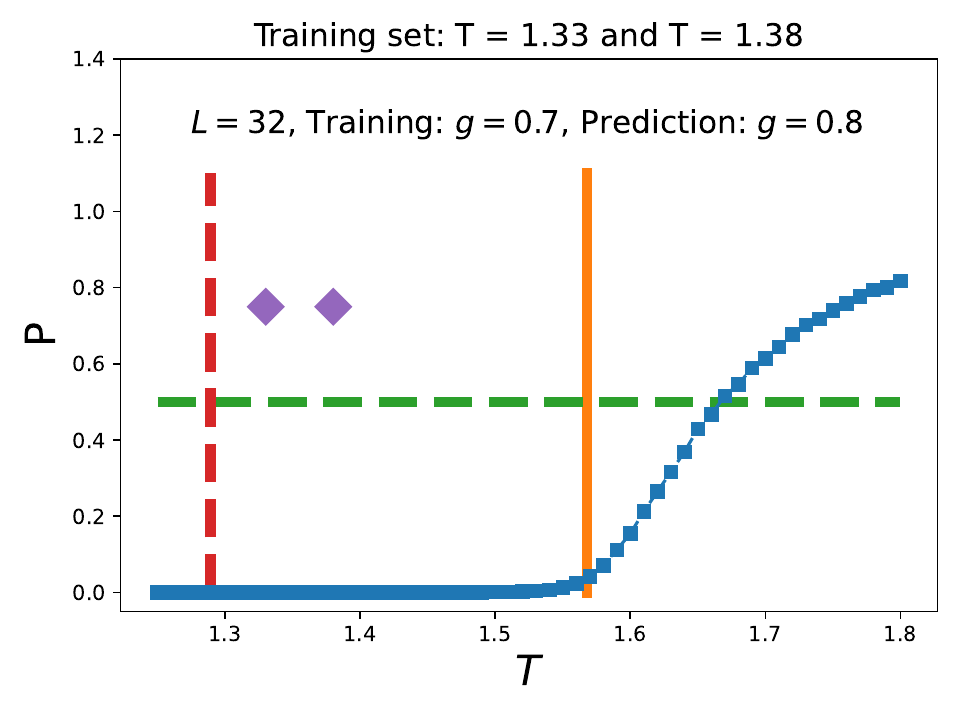}
	}
	\caption{The transfer learning results. The CNN1 outputs as functions of $T$. The horizontal dashed lines are 0.5. Diamond symbols represent the temperatures for which the associated configurations are considered in the training sets. The vertical dashed and solid lines are the theoretical $T_c$ of $g=0.7$ and $g=0.8$, respectively. These outcomes are associated with CNN1.}
	\label{cnn04}
\end{figure}

Similarly, when both the chosen temperatures are below or are above the $T_c$ of $g=0.7$, the obtained CNN1 cannot calculate the $T_c$ of $g=0.8$ with high accuracy or the estimated $T_c$ receives a very large finite-size effect, see both panels of fig.~\ref{cnn04}

\begin{figure}
	\hbox{
	\includegraphics[width=0.45\textwidth]{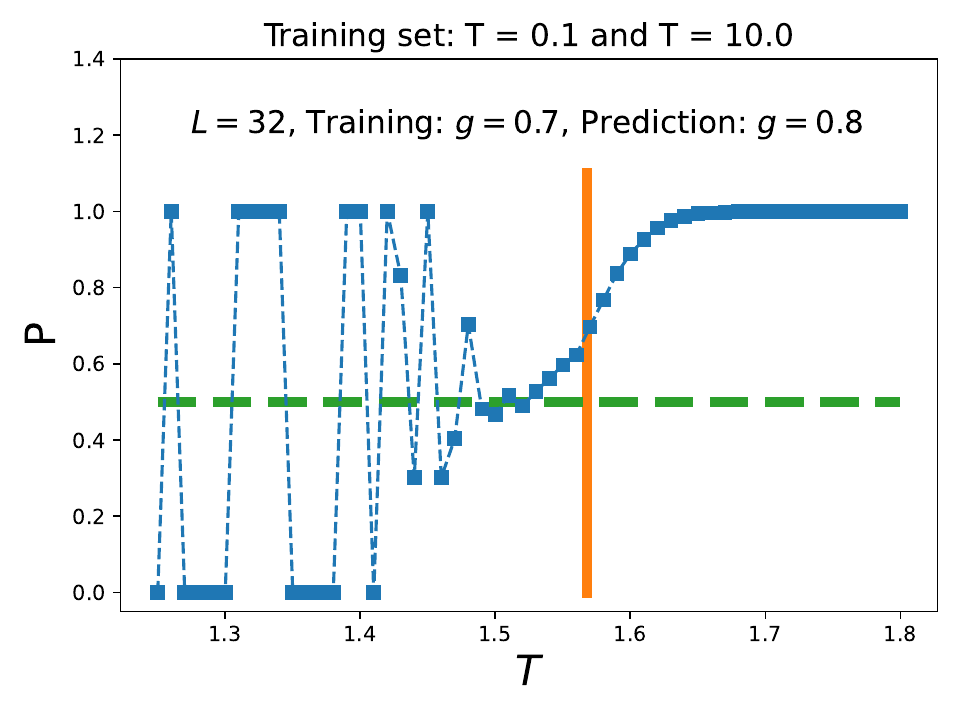}
\includegraphics[width=0.45\textwidth]{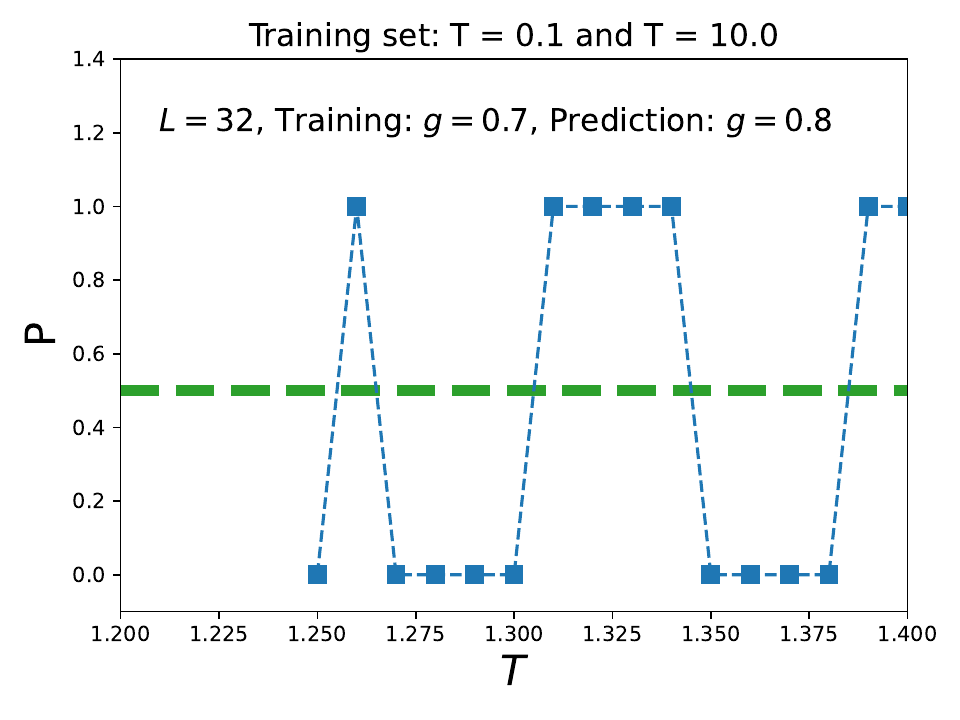}
}
	\caption{The transfer learning results. (Left) The CNN1 outputs as a function of $T$. The horizontal line is 0.5. The training set consists of configurations of $T=0.1$ and $T=10.0$. The vertical solid line is the theoretical $T_c$ of $g=0.8$. (Right) The zoom in results of the left panel.}
	\label{cnn05}
\end{figure}

Fig.~\ref{cnn05} shows the CNN1 outputs as a function of $T$ when the training set
only contains the configurations of $T=0.1$ and $T=10.0$ of $g=0.7$. The outcomes of the figure indicate the outputs jump between 1 and 0 at the low-$T$ region. This phenomenon is also found in figs.~\ref{cnn03} and \ref{cnn04}.

\begin{figure}
	\hbox{                           
		\includegraphics[width=0.33\textwidth]{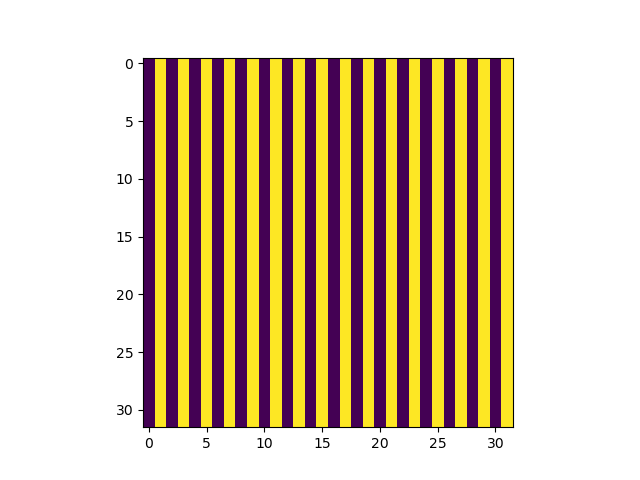}
		\includegraphics[width=0.33\textwidth]{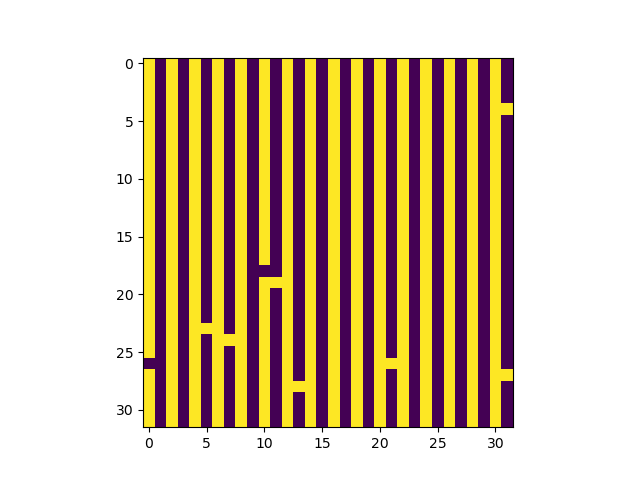}
		\includegraphics[width=0.33\textwidth]{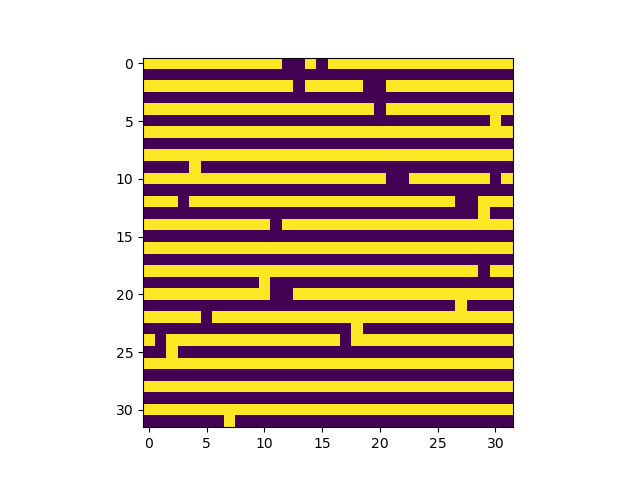}	}                    
	\caption{Typical snapshots of spin configurations for $T=0.1$ of $g=0.7$ (left), $T=1.25$ of $g=0.8$ (middle), and $T=1.26$ of $g=0.8$ (right). These outcomes are associated with CNN1.}
	\label{cnn06}
\end{figure}

The typical snapshots of spin configurations for $T = 0.1$ of $g=0.7$, $T=1.25$ of $g=0.8$, and $T=1.26$ of $g=0.8$ are depicted as the left, the middle and the right panels of Fig.~\ref{cnn06}. Clearly, one observes high similarity between
the left and the middle panels, hence leads to a CNN1 output around 0.
On the other hand, the left and the right panels are in strong contrast to each other, and one expects the CNN1 output to be around 1. In other words, the results at the low-$T$ region shown in fig.~\ref{cnn05} (see the right panel of this figure) agree well with the expectations from the snapshots of fig.~\ref{cnn06}. In particular, the phenomenon of CNN1 outputs jumping between 1 and 0 can be understood well from the snapshots of the spin configurations. Hence, the inclusion of the configurations at low-$T$ region in the training set is thus legitimate, and the outputs from the resulting CNN1 are fully consistent with the associated snapshots of the underlying spin configurations. 

Since in this situation the P curve jumps between 0 and 1 in the low-temperature region and becomes stable for those $T$ that are larger than a particular value of temperature $T_p$, one may take the $T_p$ as the pseudo-critical temperature.

One notices that the determined values of $T_p$ in the two most right panels of figs.~\ref{cnn03}, the left panel of fig.~\ref{cnn04}, and fig.~\ref{cnn05} are far away from the expected $T_c$, implying less accurate determination of $T_c$ compared to that of fig.~\ref{cnn01}. 

\subsection{The outcomes related to CNN2}

\begin{figure}
	\hbox{                           
		\includegraphics[width=0.5\textwidth]{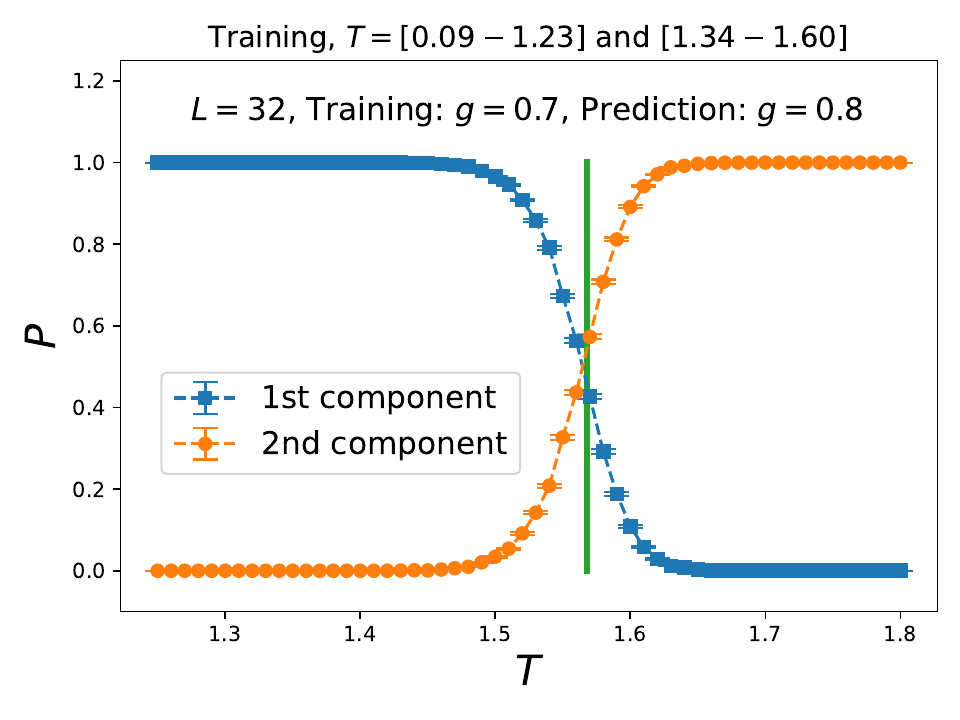}
		\includegraphics[width=0.5\textwidth]{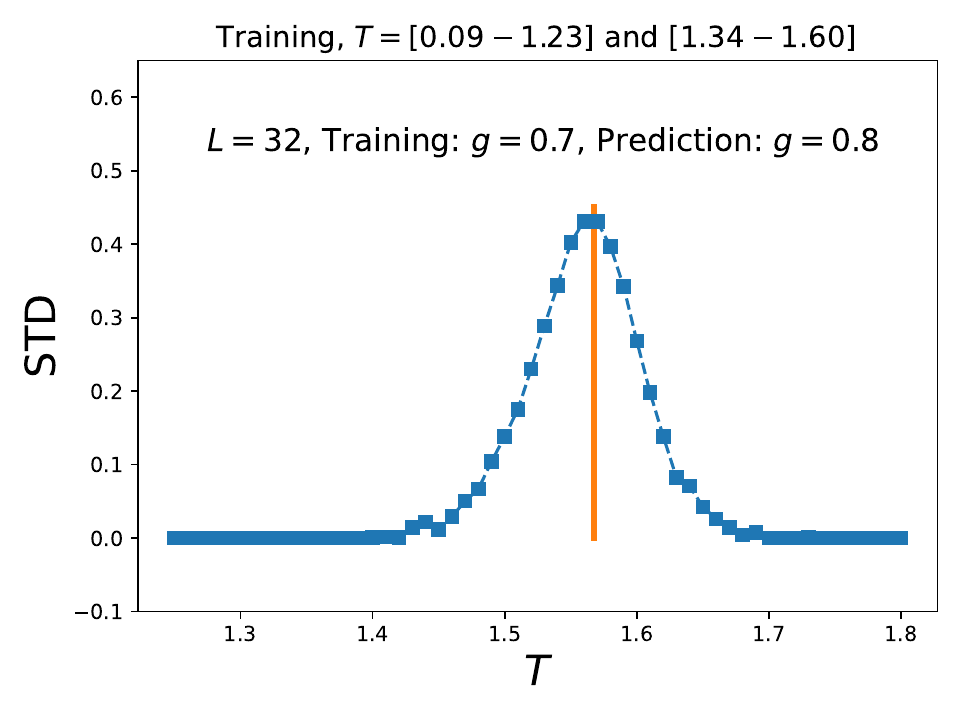}
	}                    
	
	\caption{The transfer learning results. 
		(Left) The first and the second components of the CNN2 outputs as functions of $T$. (Right) The associated standard deviations (STD) of the first component (of the CNN2 outputs) as a function of $T$.
		The vertical solid lines in both panels are the theoretical $T_c$ of $g=0.8$.}
		
	\label{cnn0}
\end{figure}

The left panel of fig.~\ref{cnn0} shows the two components of
the CNN2 outputs as functions of $T$. At what temperatures is the
training carried out is shown as the title of the panel. The title of the panel shows that the temperatures near the $T_c$ of
$g=0.7$ are excluded from being considered in the training set.
Finally, the vertical line is the expected
$T_c$ of $g = 0.8$. As can be seen from the panel, the temperature where the two components intersect agrees very well with the expected $T_c$ of $g=0.8$. This provides evidence to support the success of the proposed transfer learning using CNN2. 

The right panel of fig.~\ref{cnn0} demonstrates the standard deviations
(STD) of the first component of the CNN outputs as a function of $T$.
The results of the panel indicate that the temperature having the largest STD matches excellently with the $T_c$ of $g=0.8$. Hence, similar to the results related to CNN1, both the intersection and STD associated with CNN2 can be considered as
indicators to determine the critical temperature.

Finally, based on the CNN2 outcomes, shown in fig.~\ref{cnn0}, it is evident that the inclusion of the configurations of low-$T$ region
of $g=0.7$ in the training set(s) indeed can lead to a accurate determination of
the $T_c$ of $g=0.8$.

\begin{figure}
	\hbox{                           
		\includegraphics[width=0.5\textwidth]{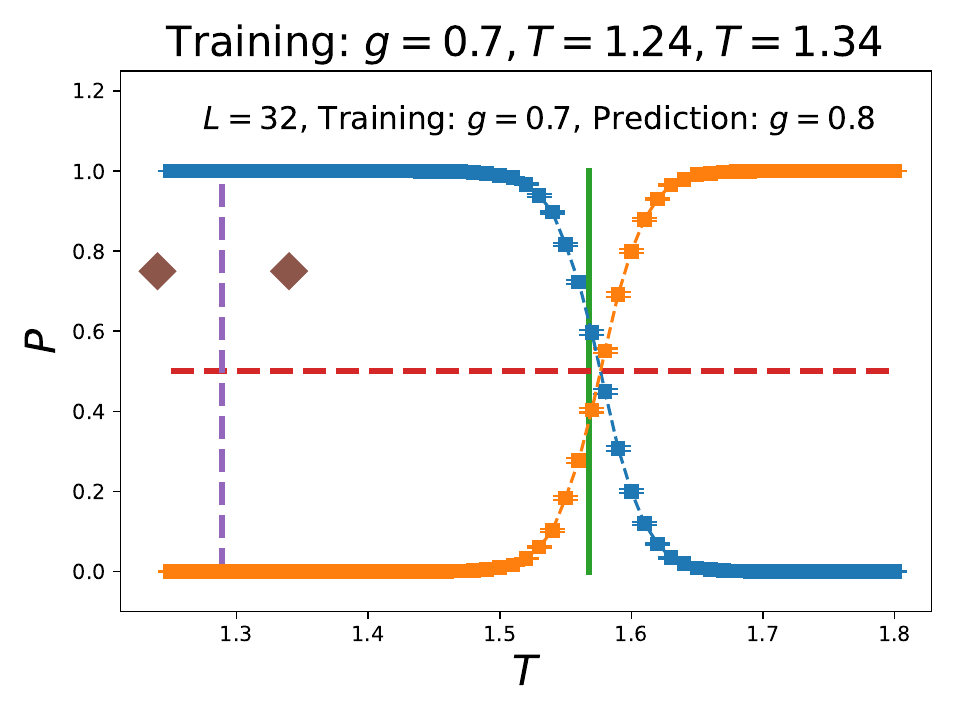}
		\includegraphics[width=0.5\textwidth]{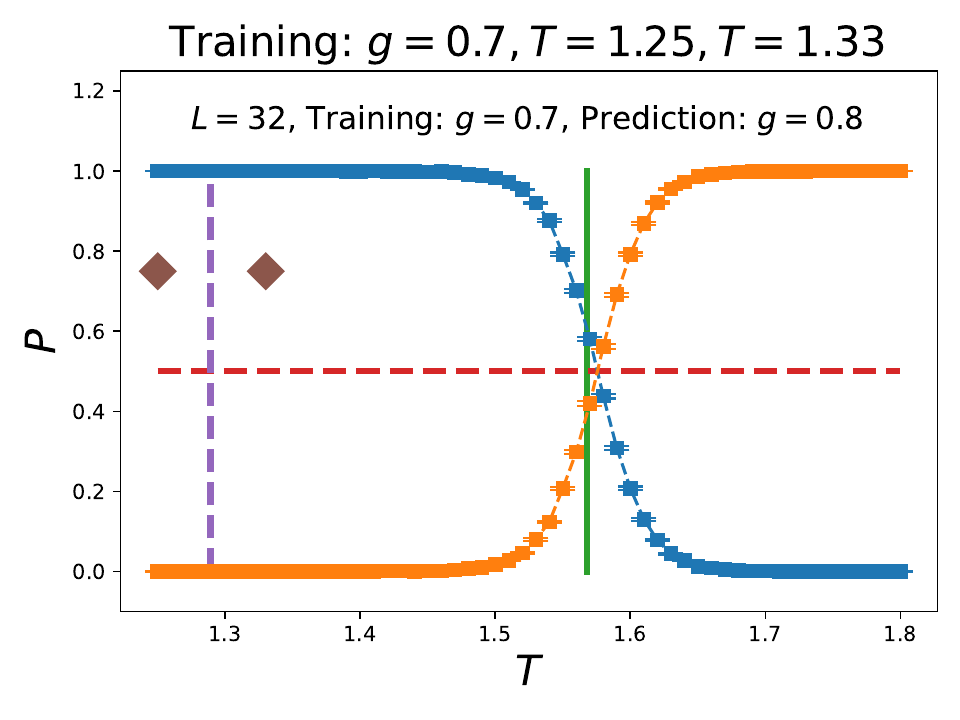}
	}                    
	
	\caption{The transfer learning results of using
		configurations at two temperatures as the training set. The diamond symbols, the dashed vertical lines, and the solid vertical lines represent the temperatures associated with training, $T_c$ of $g = 0.7$, and $T_c$ of $g=0.8$, respectively. These outcomes are associate with CNN2.}
	\label{cnn1}
\end{figure}

\begin{figure}
	\hbox{                           
		\includegraphics[width=0.5\textwidth]{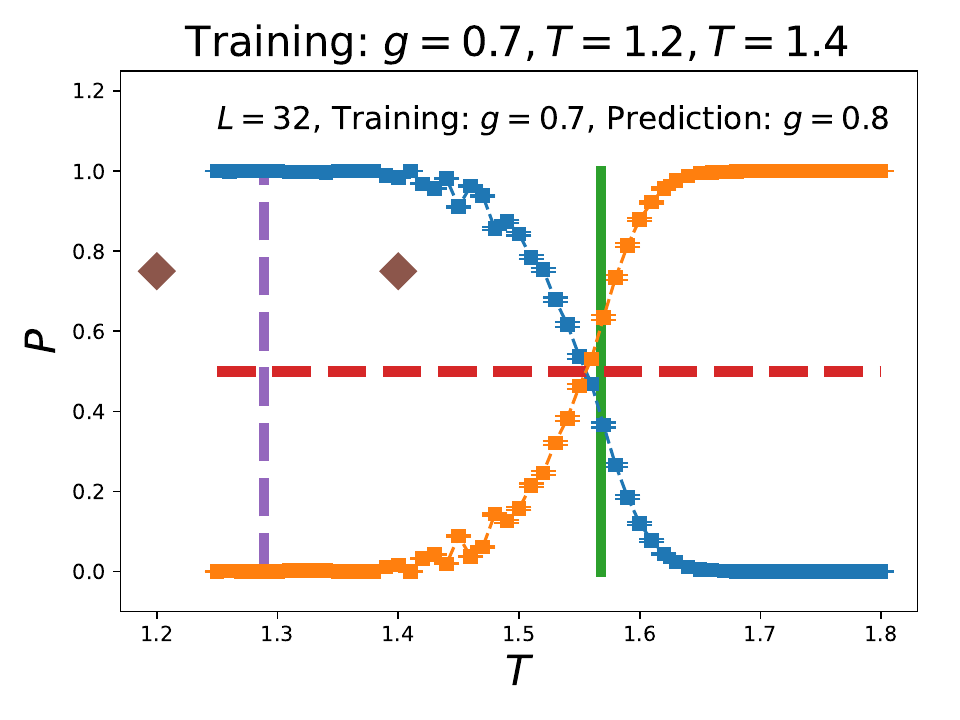}
		\includegraphics[width=0.5\textwidth]{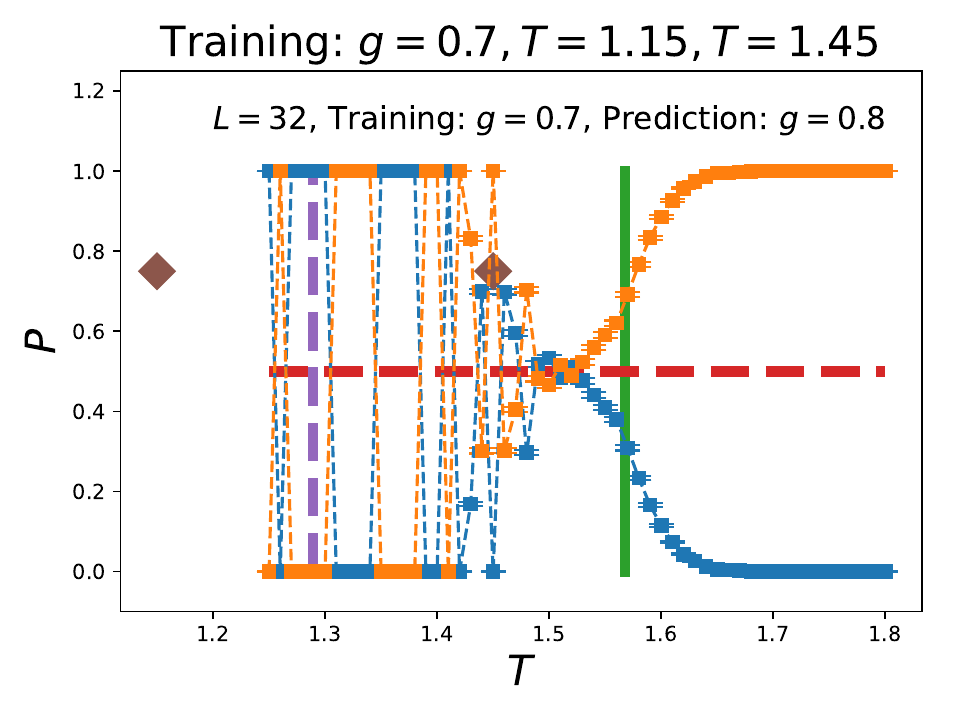}
	}                    
	
	\caption{The transfer learning results of using
		configurations at two temperatures as the training set. The diamond symbols, the bold dashed vertical lines, and the bold solid vertical lines represent the temperatures associated with training, $T_c$ of $g = 0.7$, and $T_c$ of $g=0.8$, respectively. The bold horizontal dashed line is 0.5. These outcomes are associated with CNN2.}
	\label{cnn2}
\end{figure}

Interestingly, we do find that only configurations at two temperatures
are required to train a successful CNN2. Both panels in fig.~\ref{cnn1} and the left panel of fig.~\ref{cnn2} show
the CNN2 outputs as functions of $T$ for $L=32$ and $g = 0.8$. The diamond
symbols in these panels represent the temperatures
for which the associated configurations are used as the training set.
The vertical bold dashed and bold solid lines stand for the values of $T_c$
for $g=0.7$ and $g=0.8$, respectively, and the bold horizontal dashed line is 0.5. It is clear that when the CNN2 is trained with configurations of merely two temperatures, one is above and one is below the $T_c$ of $g=0.7$ and those temperatures are not too far away from the $T_c$ of $g=0.7$, the intersection 
of the two components of the CNN2 output vectors for $g=0.8$ obtained from the trained CNN2 matches well with the $T_c$ of $g=0.8$.

\begin{figure}
	\hbox{                           
		\includegraphics[width=0.45\textwidth]{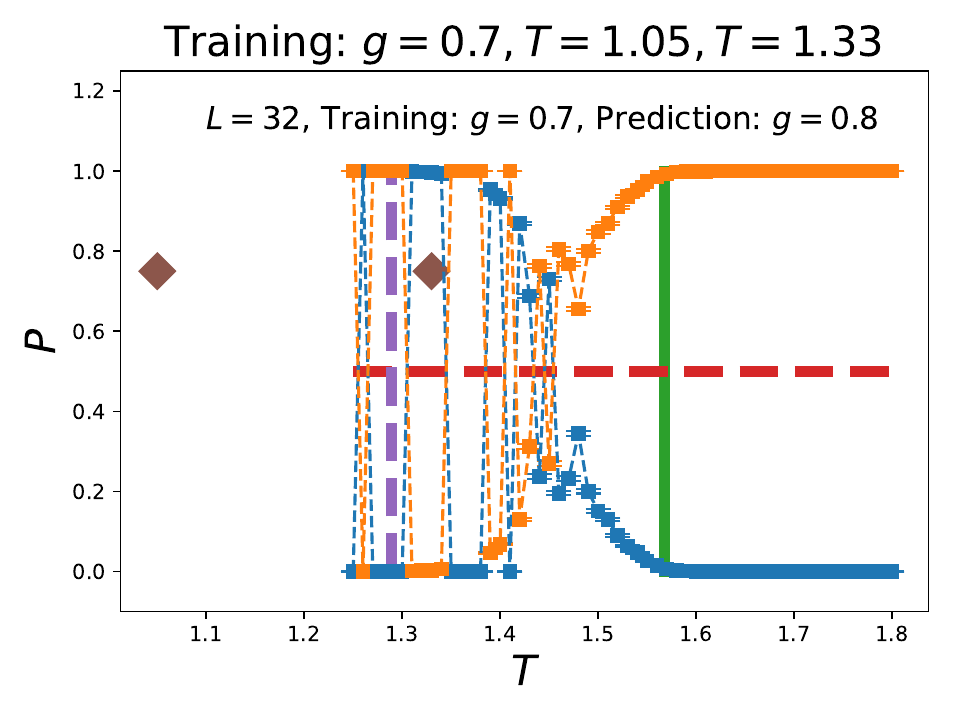}
		\includegraphics[width=0.45\textwidth]{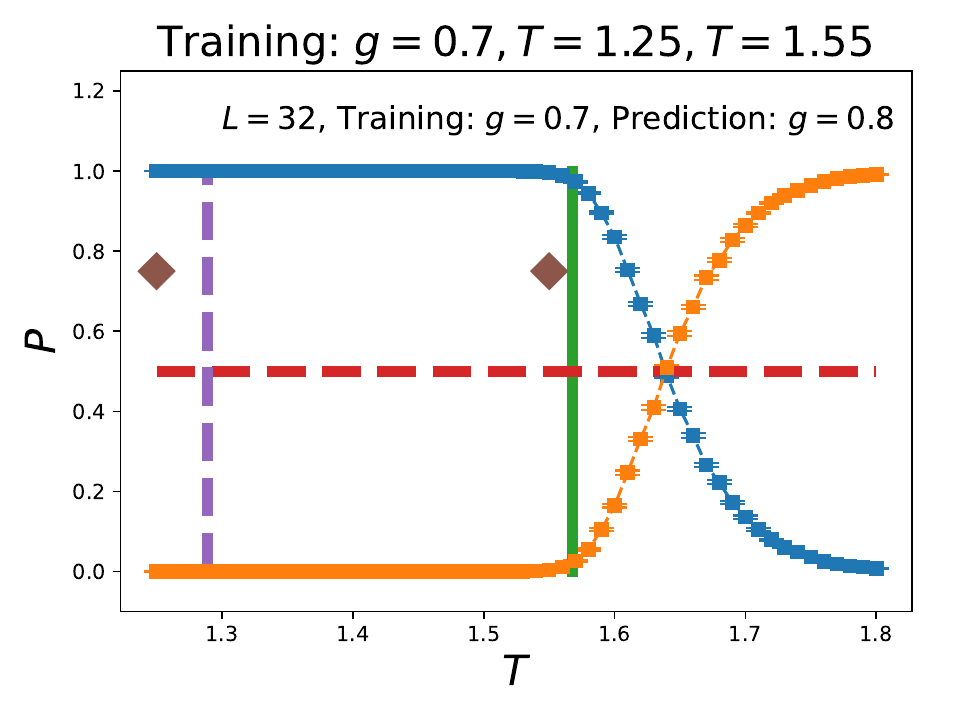}
	}                    
	
	\caption{The transfer learning results of using
		configurations at two temperatures as the training set. The diamond symbols, the bold dashed vertical lines, and the bold solid vertical lines represent the temperatures associated with training, $T_c$ of $g = 0.7$, and $T_c$ of $g=0.8$, respectively. The bold horizontal dashed line in 0.5. These outcomes are associated with CNN2}
	\label{cnn4}
\end{figure}

With the same notations and conventions introduced in the previous paragraph, the right panel of fig.~\ref{cnn2}, both panels of figs.~\ref{cnn4} suggest that
if any of the chosen two temperatures, for which the associated configurations are used as the training set, is too far away from
the $T_c$ of $g=0.7$, then the trained CNN2 either fails to determine
the $T_c$ of $g=0.8$ accurately (the left panel of fig.~\ref{cnn4}) or the NN-calculated value of $T_c$ is far away from the theoretical one (the right panel of fig.~\ref{cnn4}). If in the right panel of fig.~\ref{cnn2} and in the right panel of fig.~\ref{cnn4}, one takes the most right intersecting point associated with the two components of the CNN2 outputs as the NN-determined $T_c$, then clearly
the NN-determined values of $T_c$ are away from the theoretical $T_c$ and are less accurate than those estimated in figs.~\ref{cnn0} and \ref{cnn1}.

\begin{figure}
	\hbox{                           
		\includegraphics[width=0.5\textwidth]{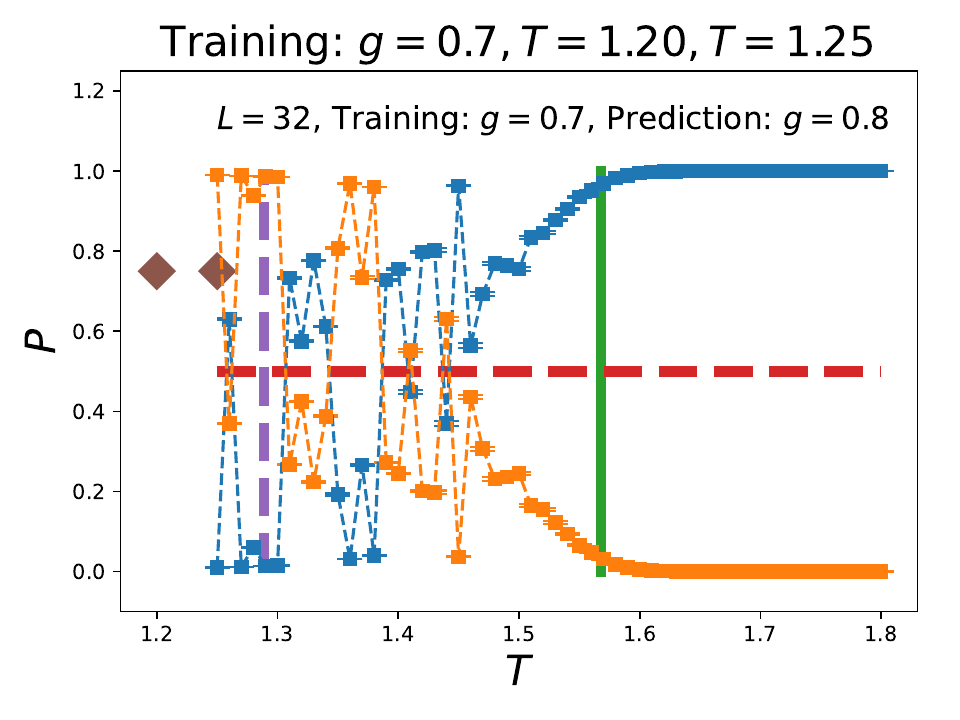}
		\includegraphics[width=0.5\textwidth]{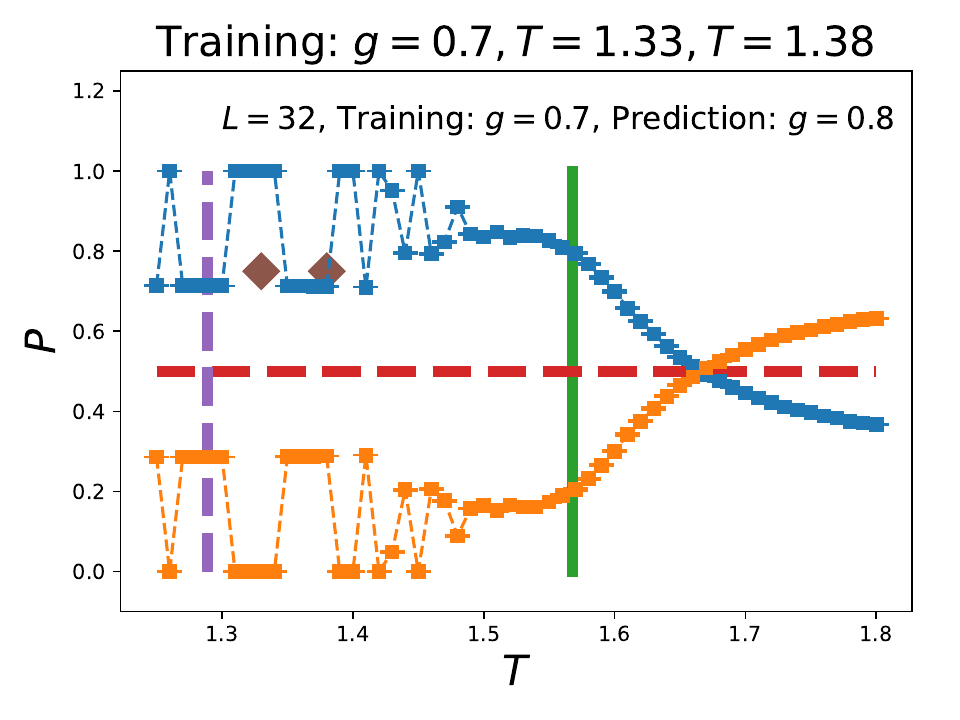}
	}                    
	
	\caption{The transfer learning results of using
		configurations at two temperatures as the training set. The diamond symbols, the bold dashed vertical lines, and the bold solid vertical lines represent the temperatures associated with training, $T_c$ of $g = 0.7$, and $T_c$ of $g=0.8$, respectively. The bold horizontal dashed line is 0.5. These outcomes are associated with CNN2.}
	\label{cnn6}
\end{figure}

Similarly, if the chosen two temperatures are both below or are both above the $T_c$
of $g=0.7$, then the resulting CNN2 either fails to determine
the $T_c$ of $g=0.8$ accurately or the NN-calculated $T_c$ is far away from the theoretical $T_c$, see fig.~\ref{cnn6}

\begin{figure}
	\hbox{                           
		\includegraphics[width=0.33\textwidth]{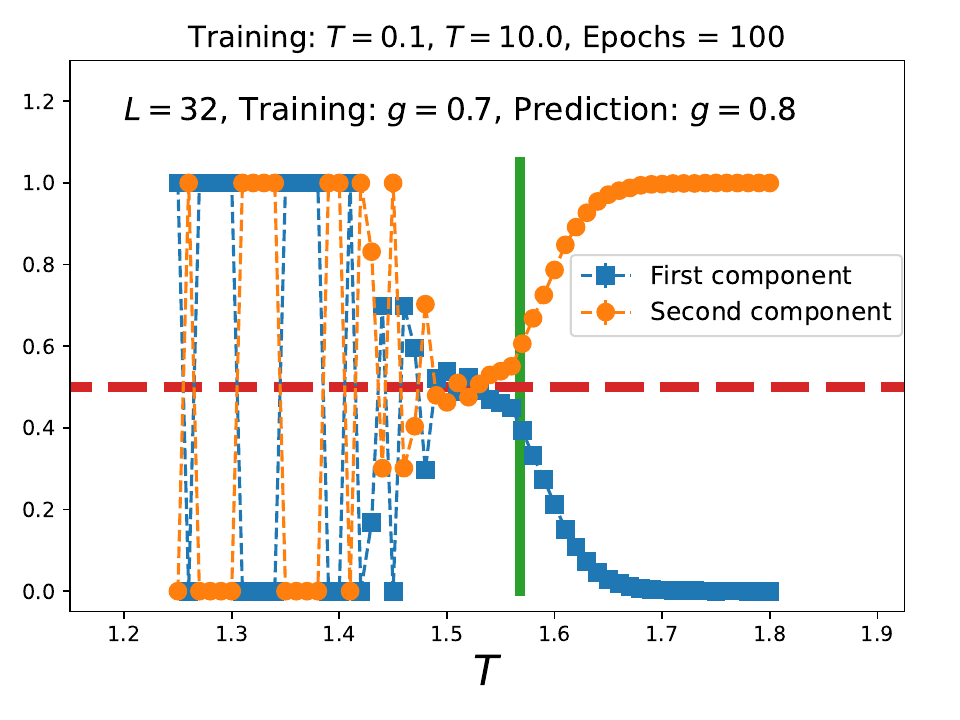}
		\includegraphics[width=0.33\textwidth]{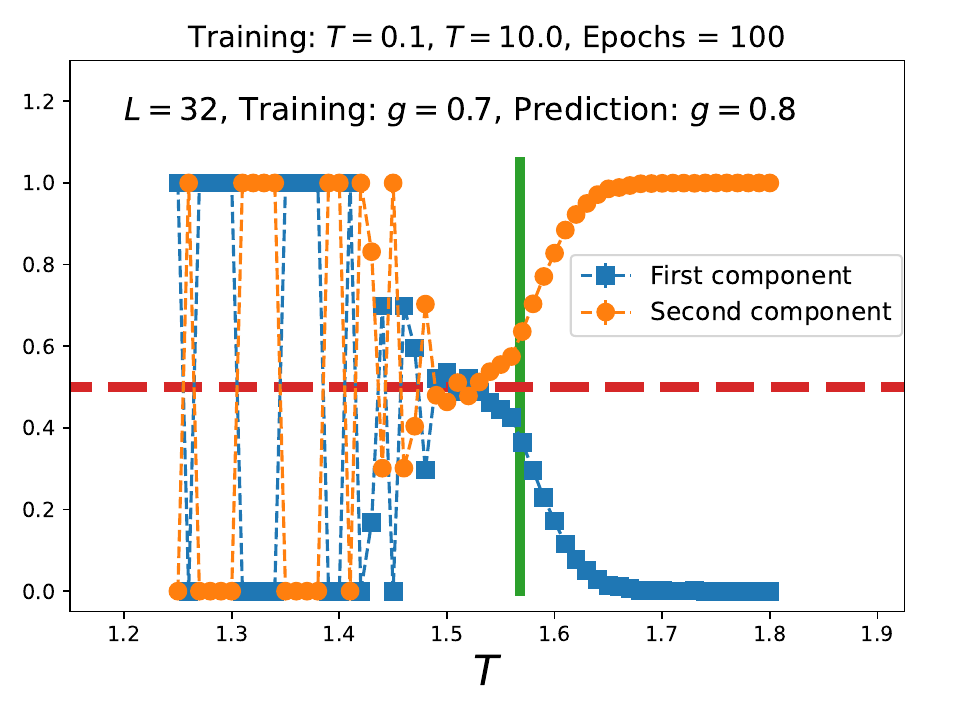}
		\includegraphics[width=0.33\textwidth]{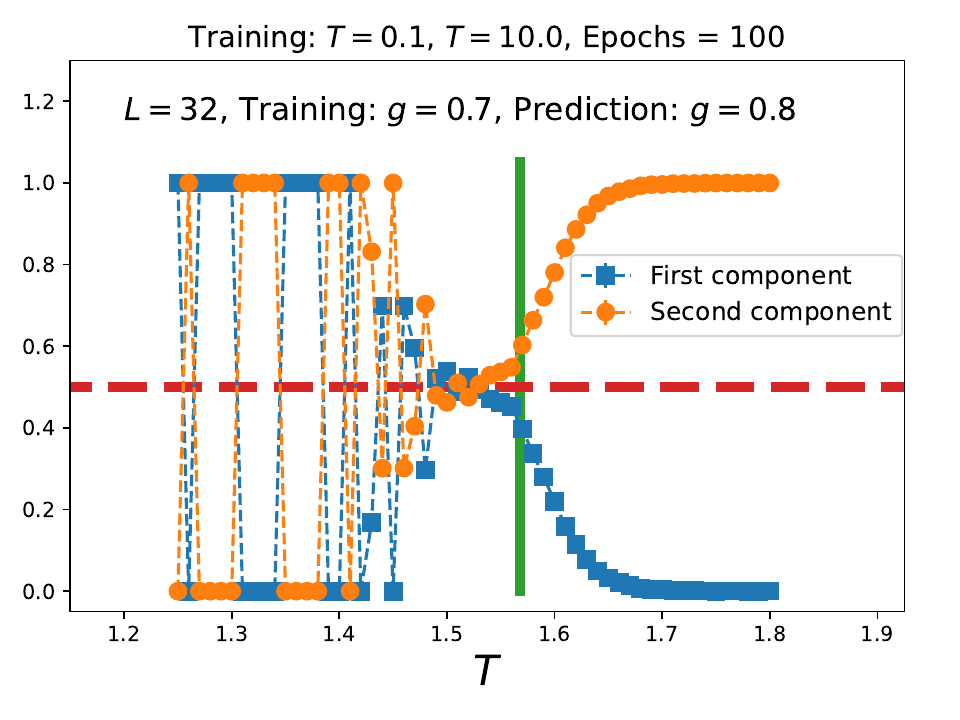}		
	}                    
	
	\caption{The transfer learning results of using
		configurations at two temperatures, namely $T=0.1$ and $T=10.0$, as the training set. The bold solid vertical line and the bold horizontal lines represent $T_c$ of $g=0.8$ and 0.5, respectively. The number of epochs considered for obtaining these results is 100 and each panel is determined with the CNN2 trained by using different sets of random seeds.}
	\label{cnn61}
\end{figure}

Finally, the CNN2 outputs as functions of $T$ obtained by conducting 100 epochs for the training and using 3 different sets of random seeds are depicted in fig.~\ref{cnn61}. The training set consists of the configurations of $T=0.1$ and $T=10.0$ of $g=0.7$. The outcomes of fig.~\ref{cnn61} reveal the same message as 
those show previously. In particular, if the temperature where both the P curves
become stable is treated as the NN-determined pseudo-critical temperature, 
then one finds that while the NN-calculated values of $T_c$ for $g=0.8$ are close to the theoretical one, they are less accurate than those shown in figs.~\ref{cnn0}, \ref{cnn1} and \ref{cnn2}.

It is also important to point out that the fluctuations appeared in the low-$T$
region in some of the figures can be explained well by the snapshots of fig.~\ref{cnn06}. In other words, the inclusion of the low-$T$ configurations in the training set is legitimate and can indeed capture the characteristics of states in the testing set(s).
 
Previously, the outcomes from CNN2 are for $L=32$. In the following, certain results associated with $L=128$ will be demonstrated.

\begin{figure}
	\hbox{                           
		\includegraphics[width=0.5\textwidth]{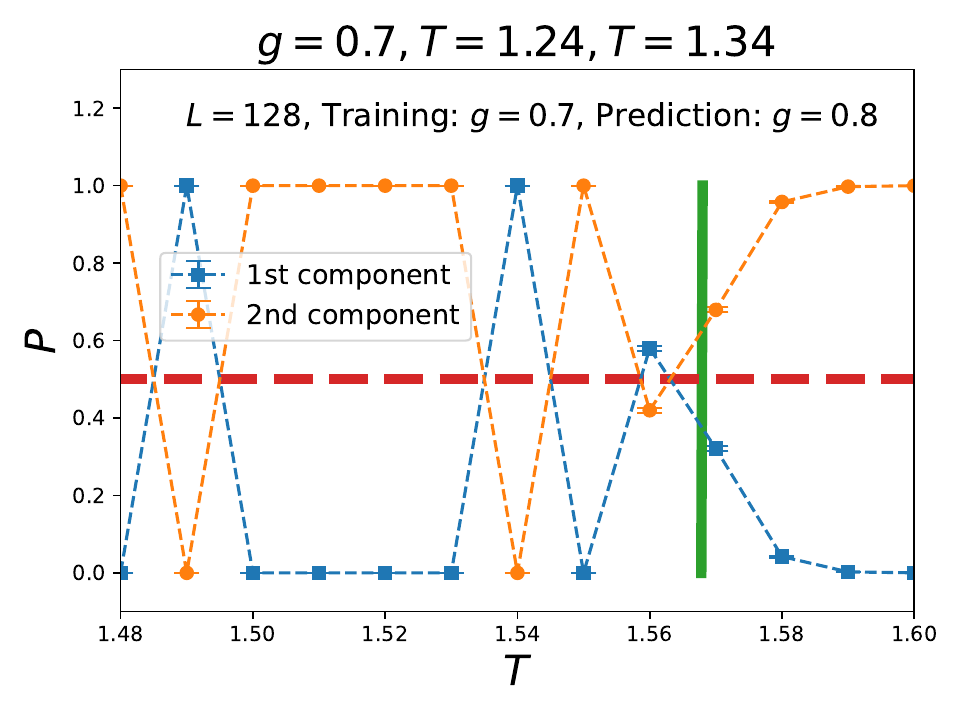}
		\includegraphics[width=0.5\textwidth]{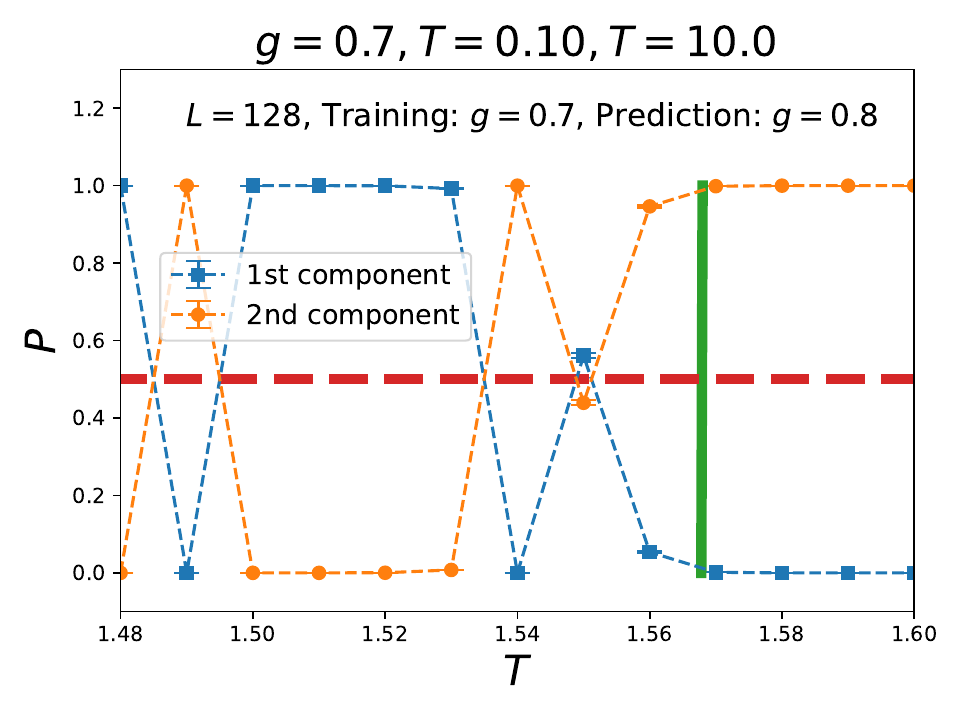}
	}                    
	
	\caption{The transfer learning results of using
		configurations at two temperatures, namely $T=1.24$ and $T=1.34$ (left panel), as the training set. The separation between two consecutive temperatures in the testing set (denoted by $\Delta T$) is 0.01. The bold solid vertical line and the bold horizontal dashed lines represent $T_c$ of $g=0.8$ and 0.5, respectively. The number of epochs considered for obtaining these results is 100. The outcomes are obtained using CNN2. The NN-related parameters for outcomes of the right panel are the same as those of the left panel, except that the two temperatures related to the training are 0.1 and 10.0.}
	\label{cnn7}
\end{figure}

\begin{figure}
	\hbox{                           
		\includegraphics[width=0.35\textwidth]{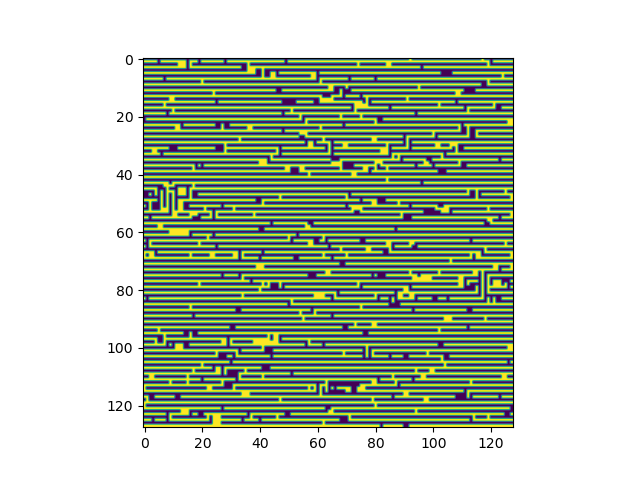}
		\includegraphics[width=0.35\textwidth]{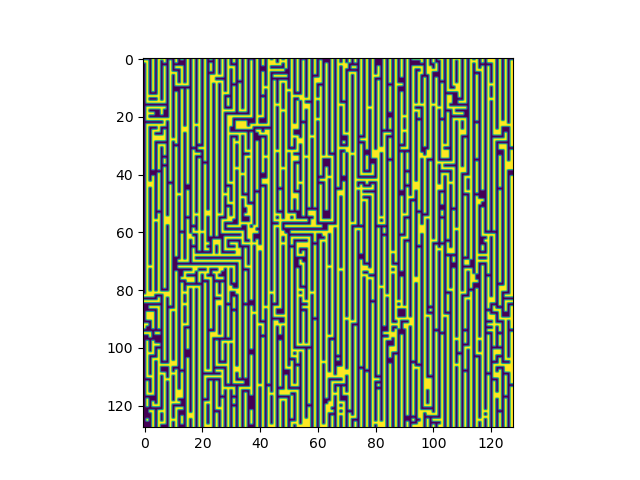}
		\includegraphics[width=0.35\textwidth]{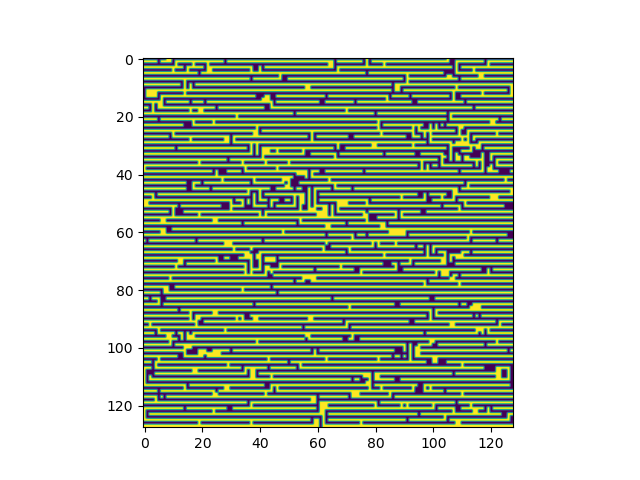}	}                    
	\caption{Typical snapshots of spin configurations for $T=1.24$ of $g=0.7$ (left), $T=1.52$ of $g=0.8$ (middle), and $T=1.49$ of $g=0.8$ (right).}
	\label{cnn71}
\end{figure}

The left and the right panels of fig.~\ref{cnn7} are the NN results for $g=0.8$ with $L=128$ and are obtained using CNN2. Both the training processes consider configurations of two temperatures (of $g=0.7$) and the used temperatures are listed as the titles of the panels. If the most right intersecting points are treated as the CNN-determined pseudo-critical temperatures, then beyond doubt the training with 
$T=1.24$ and $T=1.34$ receives (much) less finite-size effect than that of the training with $T=0.1$ and $T=10.0$. Here again, the fluctuations between 0 and 1 for the two components of the NN outputs below $T_c$ are due to the same reason as
that explaining the fluctuations shown up in figs.~\ref{cnn03}, \ref{cnn04}, and \ref{cnn05}, sees fig.~\ref{cnn71} and the associated caption.

Indeed, the similarity between the left ($T=1.24, g = 0.7$) and the right ($T=1.49, g = 0.8$) panels of fig.~\ref{cnn71} definitely will lead to a CNN2 output vector with the first component
being around 1. In addition, based on the strong contrast between the left ($T=1.24, g=0.7$) and the middle ($T=1.52, g=0.8$) panels, one would arrive at a CNN2 output vector with the first component being around 0. These arguments agrees well with the results presented in
the left panel of fig.~\ref{cnn7}.

\begin{figure}
	\hbox{                           
		\includegraphics[width=0.5\textwidth]{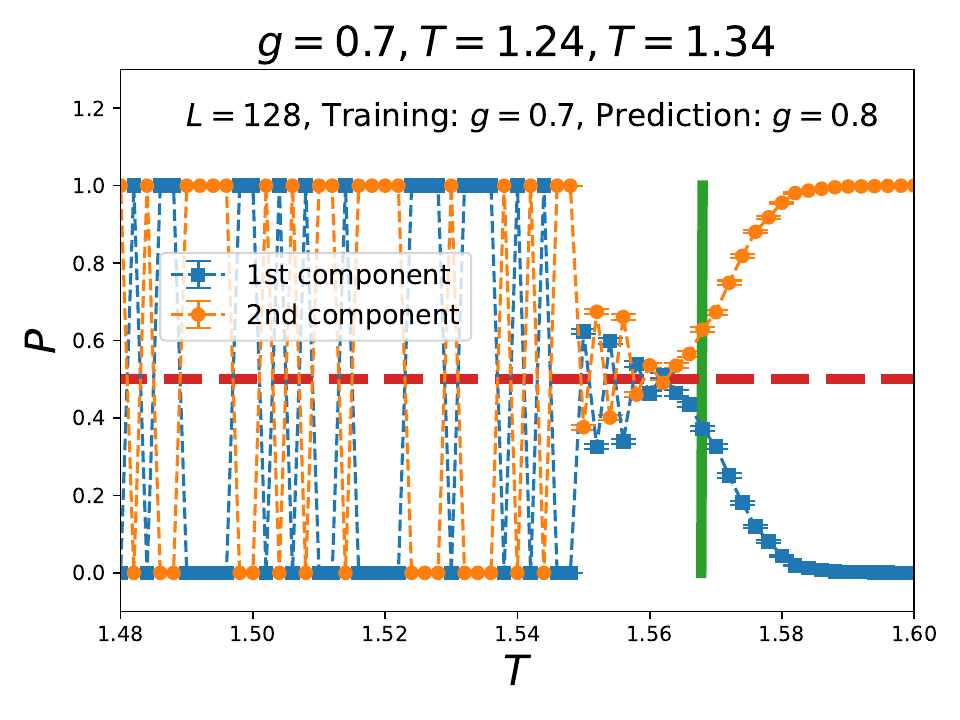}
		\includegraphics[width=0.5\textwidth]{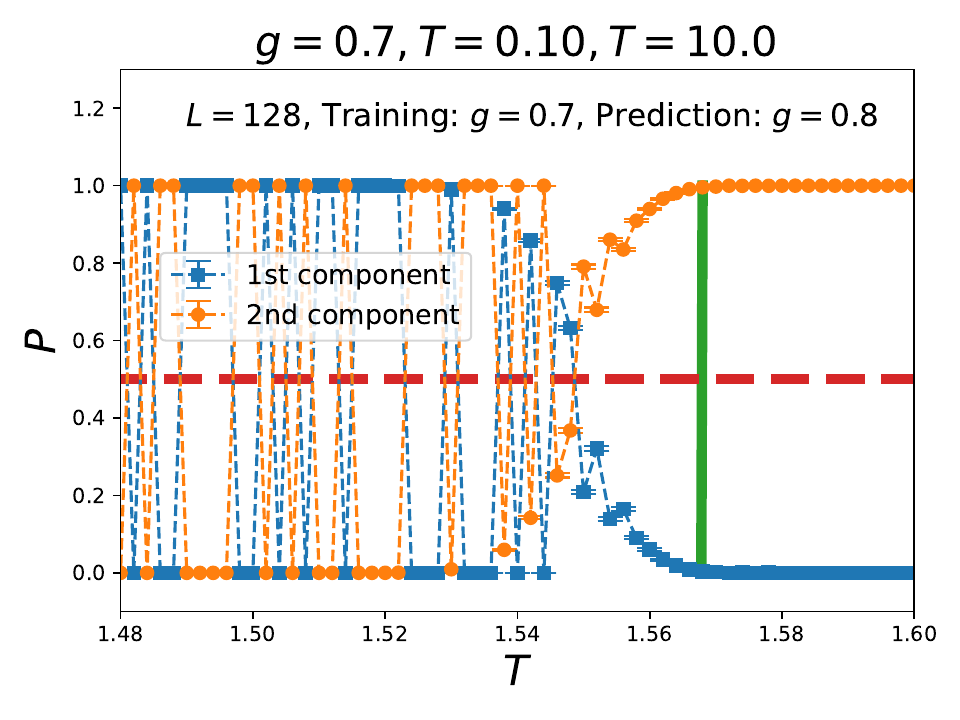}
	}                    
	
	\caption{The transfer learning results of using
		configurations at two temperatures, namely $T=1.24$ and $T=1.34$ (left panel), as the training set. $\Delta T = 0.002$. The bold solid vertical line and the bold horizontal dashed lines represent $T_c$ of $g=0.8$ and 0.5, respectively. The number of epochs considered for obtaining these results is 100. The outcomes are obtained using CNN2. The NN-related parameters for outcomes of the right panel are the same as those of the left panel, except that the two temperatures related to the training are 0.1 and 10.0.
		The outcomes of both panels are obtained using CNN2.}
	\label{cnn8}
\end{figure}

The separation between two consecutive temperatures in the testing set resulting in fig.~\ref{cnn7} is $\Delta T = 0.01$. If one considers smaller $\Delta T$, namely $\Delta T = 0.002$, then the results from CNN2 are depicted in fig.~\ref{cnn8}. Similarly, if the most right intersecting points are treated as the CNN-determined pseudo-critical temperatures, then beyond doubt the training with 
$T=1.24$ and $T=1.34$ receives less finite-size effect than that of the training with $T=0.1$ and $T=10.0$. 

\subsection{The outcomes related to CNN3}

For the last CNN3 considered here, we have focused on the linear system size $L=128$.

\begin{figure}
 	\hbox{                           
 		\includegraphics[width=0.5\textwidth]{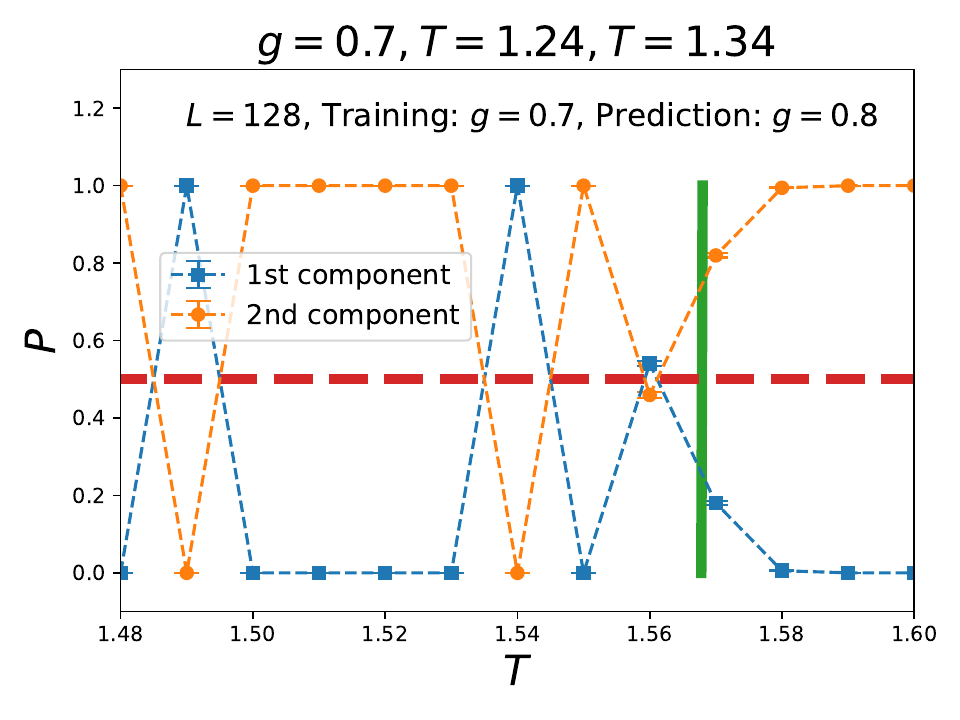}
 		\includegraphics[width=0.5\textwidth]{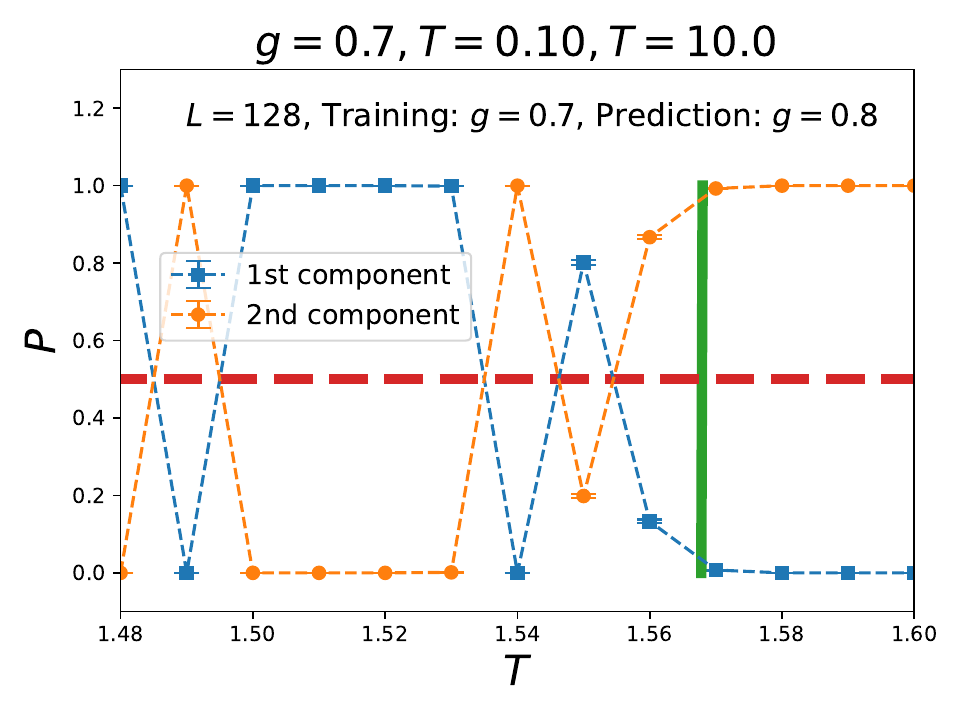}
 	}                    
 	
 	\caption{The transfer learning results of using
 		configurations at two temperatures, namely $T=1.24$ and $T=1.34$ (left panel), as the training set. $\Delta T = 0.01$. The bold solid vertical line and the bold horizontal lines represent $T_c$ of $g=0.8$ and 0.5, respectively. The number of epochs considered for obtaining these results is 100. The NN-related parameters for the outcomes of the right panel are the same as those of the left panel, except that the two temperatures related to the training are 0.1 and 10.0. The outcomes of both panels are obtained using CNN3.}
 	\label{cnn9}
 \end{figure}

For the separation between two consecutive temperatures in the testing set being $\Delta T = 0.01$, the outcomes associated with CNN3 are demonstrated in fig.~\ref{cnn9}.

The left and the right panels are the NN results for $g=0.8$ with $L=128$. Both the training processes consider configurations of two temperatures (of $g=0.7$) and the used temperatures are listed as the titles of the panels. If the most right intersecting points are treated as the CNN-determined pseudo-critical temperatures, then beyond doubt that the training with 
$T=1.24$ and $T=1.34$ receives less severe finite-size effect than that of the training with $T=0.1$ and $T=10.0$.

\begin{figure}
	\hbox{                           
		\includegraphics[width=0.5\textwidth]{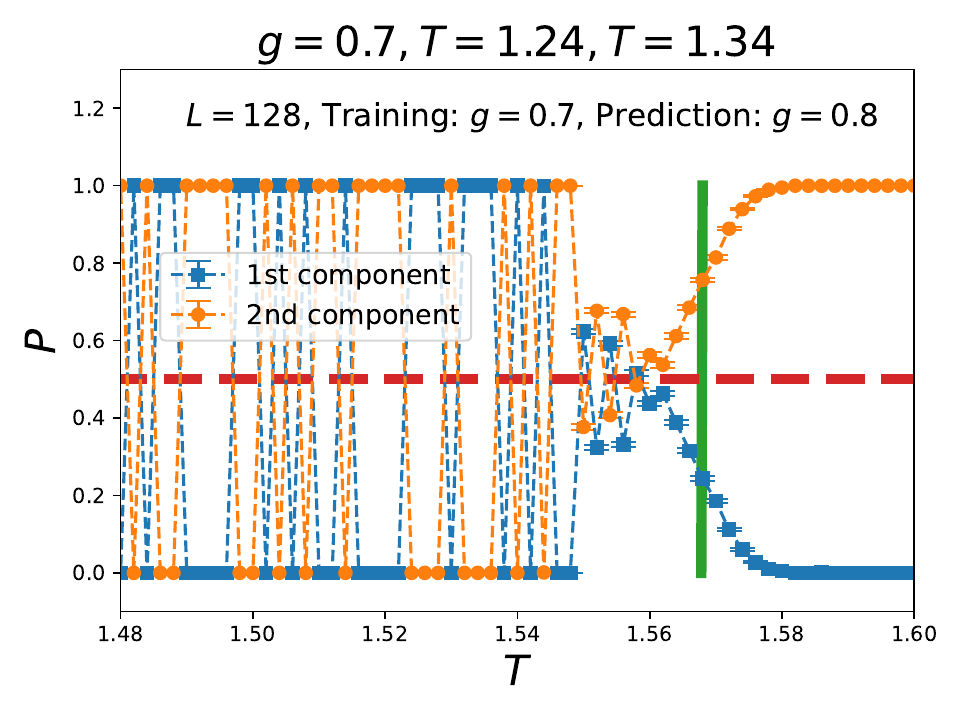}
		\includegraphics[width=0.5\textwidth]{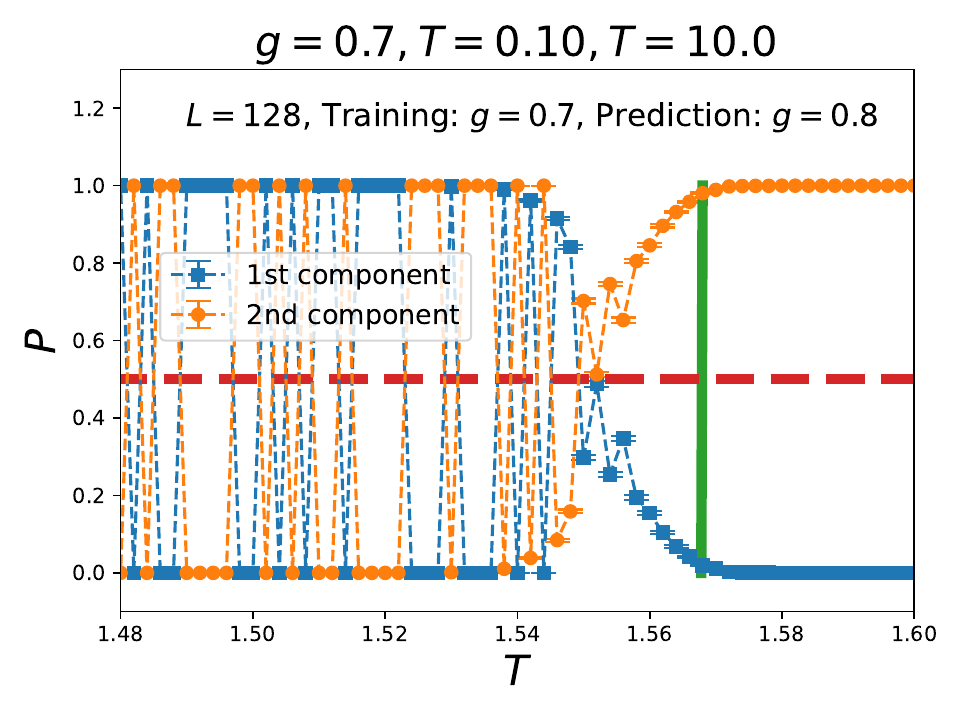}
	}                    
	
	\caption{The transfer learning results of using
		configurations at two temperatures, namely $T=1.24$ and $T=1.34$ (left panel), as the training set. $\Delta T = 0.002$. The bold solid vertical line and the bold horizontal lines represent $T_c$ of $g=0.8$ and 0.5, respectively. The number of epochs considered for obtaining these results is 100. The outcomes are obtained using CNN3. The NN-related parameters for outcomes of the right panel are the same as those of the left panel, except that the two temperatures related to the training are 0.1 and 10.0. The outcomes are obtained using CNN3.}
	\label{cnn10}
\end{figure}

If one considers the case of $\Delta T = 0.002$, then the results from CNN3 are depicted in fig.~\ref{cnn10}. Similarly, if the most right intersecting points are treated as the CNN-determined pseudo-critical temperatures, then beyond doubt the training with $T=1.24$ and $T=1.34$ receives less severe finite-size effect than that of the training with $T=0.1$ and $T=10.0$. 
 
\section{Discussions and Conclusions}

In this study, we examine the transfer learning for the frustrated $J_1$-$J_2$ Ising model on the square lattice. Specifically,
we train three CNNs using the spin configurations of $g=0.7$ as the training set. Then the obtained CNNs are employed to study the phase transition of $g=0.8$.

We find that this transfer learning is successful. In particular,
only configurations of two temperatures, one is below and one is above
the $T_c$ of $g=0.7$, are needed to train a working CNN that can determine
the critical temperature $T_c$ of $g=0.8$ precisely. 

We also find that the chosen two temperatures mentioned in the previous paragraph, which are used for the training, cannot be too far away from the $T_c$ of $g=0.7$, otherwise, the obtain CNN cannot calculate
the wanted $T_c$ of $g=0.8$ with reasonably high precision. Specially, a large finite-size effect is observed. 

In addition, despite the fact that the configurations at low-$T$ region do not contain all the possible ground state
configurations, we find they can still be considered in the training and prediction
stages, with the price of the appearance of large finite-size effects.

The model considered here have been studied in Refs.~\cite{Cor21,Civ25} using a supervised CNN (with transfer learning) and an unsupervised variational auto-encoder. Configurations at multiple temperatures, both (far) below and above the $T_c$, are considered as the training set(s) in these two studies. As a result, it is likely that the training sets of \cite{Cor21,Civ25} contain biased states like ours since local spin flip algorithms are used in Refs.~\cite{Cor21,Civ25}. However, similar to our outcomes shown in figs.~\ref{cnn01} and \ref{cnn0}, satisfactory results are obtained in these two references since the role of biased configurations at low-$T$ region in the training stage are tamed by the configurations below but near $T_c$.

Because the training process is the most time-consuming procedure in a NN study of critical phenomena, it is indeed desirable to obtain a strategy of shortening the time and reducing the computing resources required to train a NN. The idea of employing as few spin configurations as possible in the training set is one route to reach this goal. In addition, using configurations from temperatures far away from the critical points makes the supervised training procedure be similar to the unsupervised ones. Our investigation presented here demonstrates that the use of real configurations (not the theoretical ones) from $T = 0$ and $T= \infty$ as the training set may seriously sacrifice the accuracy of the determined critical points (hence the critical exponents). As a result, in a NN calculation, the method of using only the real configurations at the two ends of the considered parameter as the training set may not be an ideal approach since one may not reach very high precision for the critical points and exponents with such a training strategy. 

\begin{figure}
	\vbox{
		\hbox{                           
			\includegraphics[width=0.45\textwidth]{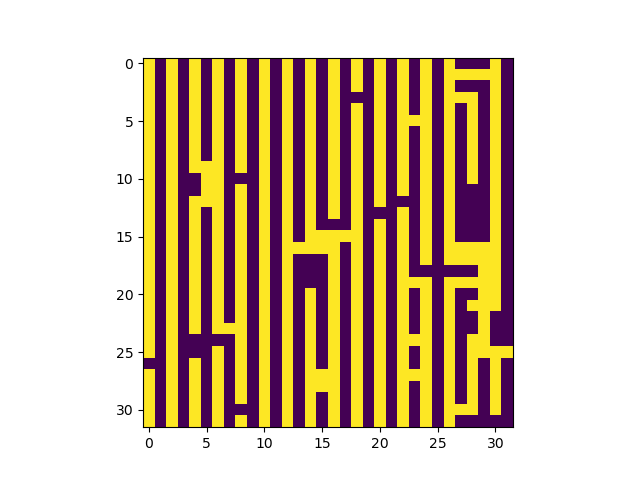}
			\includegraphics[width=0.45\textwidth]{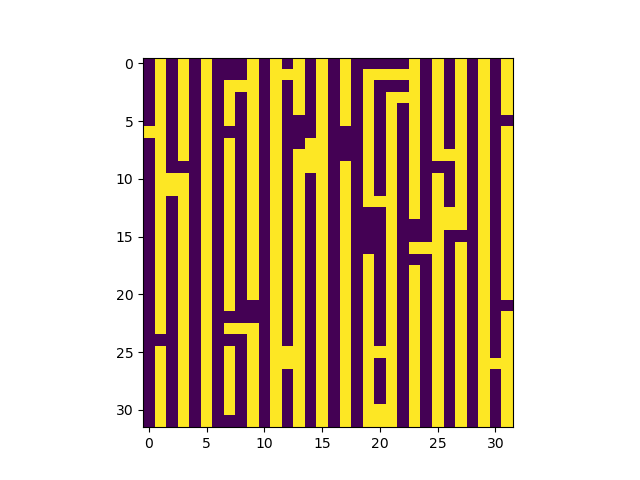
		}} 
		\hbox{                           
			\includegraphics[width=0.45\textwidth]{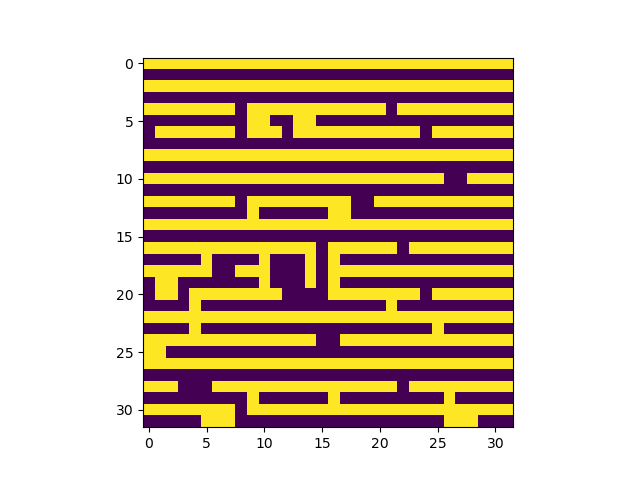}
			\includegraphics[width=0.45\textwidth]{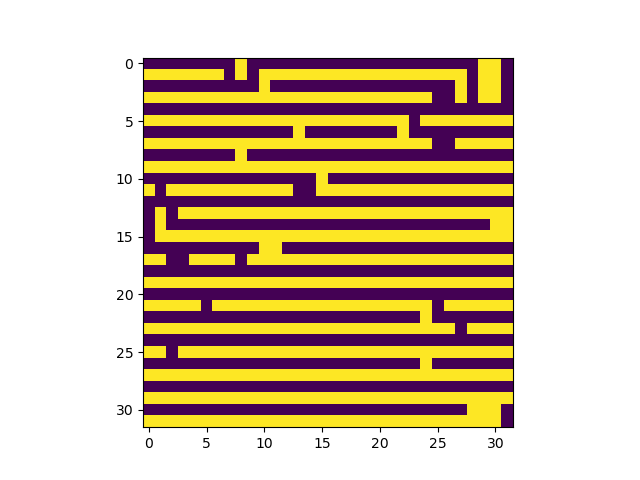}
		}                         
	}
	\caption{The snapshots of $g=0.7$ with $L=32$ and $T=1.25$.}
	\label{stripe}
\end{figure}

Of course, our conclusions are based on the fact that the MC algorithm used here is not able to sample the configurations at low-$T$ region efficiently. Hence, the observed large finite-size effect(s) and the phenomenon of the CNN outputs jumping between 0 and 1 may attribute to this. For real experiments,
it can be the case that the data at the two ends of the considered parameter may be sufficient to train a CNN successfully so that it can determine the critical points and the associated exponents with acceptable accuracy. This success crucially relies on the condition that the data at the two ends of the considered parameter contain (almost) all the information (all the possible states) of the studied system. 

To understand the success of determining the $T_c$ accurately as that shown in the right panel of fig.~\ref{cnn1} (Which uses only configurations of $T=1.25$ and $T=1.33$ as the training set), we have investigated the spin configurations of $T=1.25$. Interestingly, we do
find that the training set contains configurations similar to the four expected ground state stripe configurations, see fig.~\ref{stripe}. 
This outcome provides convincing evidence to support our claim made in the previous paragraph. Specifically, the training of a NN would be successful
if the data at the two ends of the considered parameter contain (almost) all the information (all the possible states) of the studied system. 
Finally, the left panel of fig.~\ref{cnn03} and the right panel of
fig.~\ref{cnn4} suggest that the separation between the high end parameter
and the true $T_c$ cannot be too larger than that associated with the low end parameter as well.

In conclusion, conducting the required NN calculations is needed to determine whether the idea of using the data at the two ends of the considered parameter can lead to a NN that can be used to explore the wanted
critical phenomenon accurately. In particular, it is crucial that the data at the low end of the considered parameter should contain all the possible ground-state like configurations, and the separation between one end of the studied parameter and the true $T_c$ cannot differ too much from that related to the other end of parameter, to ensure a successful training.

\section*{Funding}
Partial support from National Science and Technology Council (NSTC) of
Taiwan is acknowledged (Grant numbers: NSTC 112-2112-M-003-016- and NSTC 113-2112-M-003-014-).

\section*{Conflict of Interest}
The authors declare no conflict of interest.

\section*{Data Availability Statement}
Data are available from the corresponding author
on reasonable request.

\end{document}